# Integrated pulse scope for tunable generation and intrinsic characterization of structured femtosecond laser

Tiancheng Huo[1,2], Li Qi[1], Jason J. Chen[1,2], Yusi Miao[1,2], Yan Li[1,2], Zhikai Zhu[1,2] and Zhongping Chen[1,2*]

[1]Beckman Laser Institute and Medical Clinic, Irvine, USA.

[2]Department of Biomedical Engineering, University of California, Irvine, Irvine, USA.

*email: z2chen@uci.edu

**Abstract**

Numerous techniques have been demonstrated for effective generation of orbital angular momentum-carrying radiation, but intracavity generation of continuously tunable pulses in the femtosecond regime remains challenging. Even if such a creation was realized, the generated pulses—like all pulses in reality—are complex and transitory objects that can only be comprehensively characterized via multidimensional spaces. An integrated lasing system that generates pulses while simultaneously quantifies them can achieve adaptive pulse tailoring. Here, we report a femtosecond pulse scope that unifies vector vortex mode-locked lasing and vectorial quantification. With intracavity-controlled Pancharatnam-Berry phase modulation, continuous and ergodic generation of spirally polarized states along a broadband higher-order Poincaré sphere was realized. By intrinsically coupling a two-dimensional polarization-sensitive time-scanning interferometer to the laser, multidimensional spatiotemporal features of the pulse were further visualized. The proposed methodology paves the way for design optimization of ultrafast optics by integrating complex femtosecond pulse generation and structural customization, facilitating its applications in optical physics research and laser-based manufacturing.

## Introduction

The photon carries an orbital angular momentum (OAM) and a spin angular momentum (SAM), which are not strictly distinguishable and separately conserved[1]. Under the generally accepted paraxial approximation, the analogy between the time dependent Schrodinger equation and optical scalar wave equation[2] suggests that the linearly polarized Laguerre-Gaussian mode with null radial index will carry a well-defined OAM, which may be associated with the phase factor for an optical vortex along its axis[3]. OAM-carrying beams have recently attracted major attention in the theoretical research[4-5] and for its applications in optomechanics[6], optical communications[7], particle acceleration[8], microscopy[9-10], optical trapping and tweezing[11-12], material processing[13], remote sensing[14-15], and quantum sciences[16-17].

A range of external-cavity techniques have been developed for creating OAM-carrying beams by using spatial light modulators[18-19], Q-plates[20-21] or optical metasurfaces[22], sectored spatially varying retarder[23], and diffractive optics[24-25]. Due to insertion loss, polarization dependency, and dispersion, however, applying these techniques to a broadband source has been problematic. There even exists a paradox that modulating any freedom of a laser (e.g., polarization, spectrum, wavefront) would inevitably affect one the other. Since the recent introduction of the helical modes of light[26-29], researchers have been focusing on creating OAM-carrying pulses via an intracavity means as all spatial and spectral freedoms of the pulse intrinsically interact with each other. In an intracavity-controlled laser, mode-locking is typically employed to generate ultrashort vector vortex pulses in the picosecond or femtosecond range[30-35]. Although a variety of cavity designs have been proposed, a tunable OAM-carrying femtosecond laser with continuous and ergodic Pancharatnam-Berry phase modulation within the cavity would further allow for the generation of extremely short, refined, and high energy structured pulses.

While pulse characterization is as essential as pulse generation in ultrafast optics, comprehensive spatiotemporal quantification of a femtosecond pulse remains challenging because of its short duration and broadband nature. This is even more difficult if the pulse has a complex vectorial structure. Even though state-of-the-art streak cameras have reached a temporal resolution of ~100 fs[36], direct pulse measurement is still limited by the photodetector responding time.

Among the indirect measurement approaches, the dechirped pulse duration is commonly tested through intensity correlation[37]. Spectral analysis methods, including frequency-resolved optical gating (FROG)[38], spectral-phase interferometry for direct electric-field reconstruction (SPIDER)[39], multiphoton intrapulse interference phase scan (MIIPS)[40], and dispersion scan[41], can resolve the temporal profile and the relative spectral phases, but the spatial and polarization profile cannot be fully acquired even with recent modifications[42-48]. Nevertheless, characterization of wideband vector vortex beams has only been partially and qualitatively realized[49-50], in which the spectral phase and temporal information are lost.

In this study, we report an integrated system for intracavity generation and multidimensional characterization of broadband OAM-carrying femtosecond pulses coupled with a cylindrically symmetric polarization structure that can be mapped onto the higher-order Poincaré (HOP) sphere state (HOP$_{SS}$). We first demonstrate a novel V-shaped design for modulating Pancharatnam-Berry phase within the cavity that allows for passive mode-locking, high switching efficiency, ultra clean, and broadband output. To comprehensively evaluate the pulse quality, we then present a polarization-sensitive spatiotemporal pulse scope for characterizing the generated beams. The proposed system will provide complete knowledge of the complex pulses and be used for real-time adaptive optimization and generation of mode-locked laser.

## Results

**Intracavity-controlled OAM-carrying femtosecond laser.** Figure 1 depicts the schematic diagram of the pulse scope integrated with intracavity-controlled OAM-carrying laser and the characterization system, in which two reference systems, the global associated with the laboratory and the local with the charge-coupled device (CCD), are denoted by *oxyz* and *o'x'y'z'*, respectively. The positive direction of *oz* and *o'z'* refers to the propagation direction of the sample beam and of the reference beam, respectively. While several types of variable spiral plates (VSPs) could be used for generating OAM-carrying beams, the geometric phase elements can provide a direct connection between the SAM and OAM which serves as a spin-orbit-converter. In our design, the laser cavity is equipped with a Q-plate (q = +1/2)[20] to transform the left and right circular polarizations states ($|L\rangle$ and $|R\rangle$, respectively) to the eigenstates HOP$_{SS}$ with opposite SAM and OAM ($\pm 2q\hbar$), denoted by: $|+1, R\rangle$ and $|-1, L\rangle$, respectively.

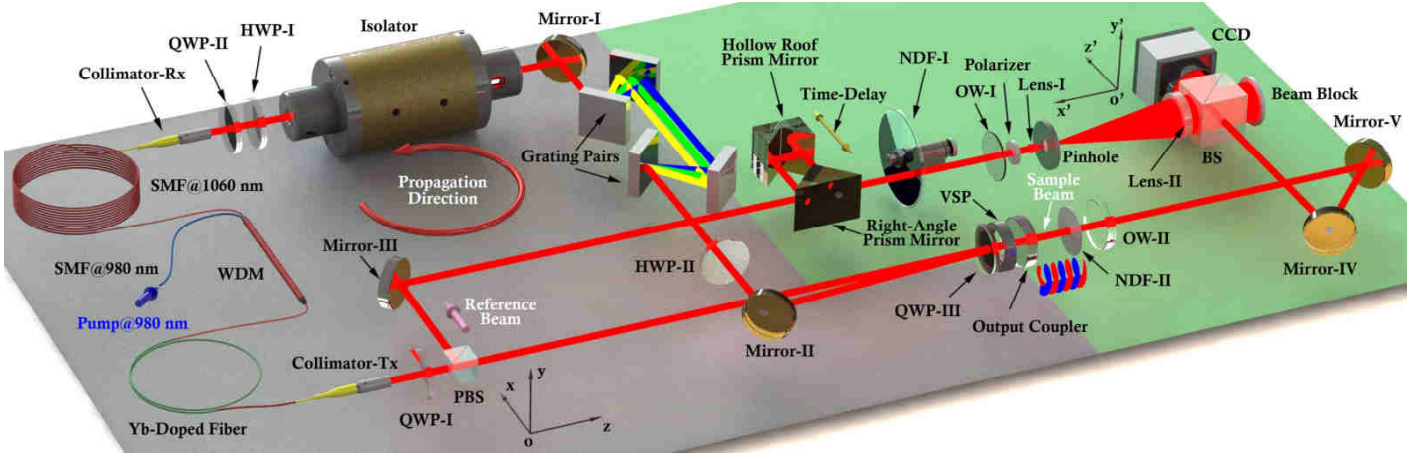

**Figure 1.** Schematic diagram of the integrated pulse scope. The gray portion of the background floor highlights the intracavity-controlled femtosecond laser, whereas the green portion of that describes the polarization-sensitive spatiotemporal characterization system. VSP = variable spiral plate; QWP = quarter wave plate; HWP = half wave plate; SMF = single mode fiber; WDM = wavelength division multiplexer; BS = beamsplitter; PBS = polarizing beamsplitter; OW = optical window; NDF = neutral density filter. The red circular arrow denotes the direction of the cavity modes.

Dispersion is an inherent dilemma in femtosecond pulses with wide spectral bandwidth. It occurs in not only the spectral and spatial phase but also the geometric phase, which results in light of different wavelengths catches different OAM and polarization distribution. Therefore, we describe the output states by using a *mean* HOP sphere ($\bar{S}^{+1}$) which corresponds to the center wavelength ($\lambda_0$). The north ($\Phi, \pi/2$) and south ($\Phi, -\pi/2$) poles represent the two eigenstates carrying spiral wavefronts with topological charge of 1 ($|(\Phi, \pi/2)\rangle = |-1, L\rangle$ and $|(\Phi, -\pi/2)\rangle = |+1, R\rangle$, respectively). In addition, the expressions for the radial and azimuthal polarization carrying states are $|(0,0)\rangle$ and $|(\pi, 0)\rangle$, respectively. By modulating the angles between the x-axis of the laboratory system (*x*) and the optical axis of quarter wave plate III (QWP-III), α, and between *x* and that of the Q-plate, β, the input horizontal polarization states are mapped onto the entire *mean* HOP sphere [$\bar{S}^{+1}(\Phi(\alpha, \beta), \Theta(\alpha, \beta))$], where $\Phi$ and $\Theta$ are the polar and azimuthal angles of the spherical coordinates, respectively. The relationship between $\Phi, \Theta$ and α, β is given by: $\Phi(\alpha, \beta) = 2(\alpha - \beta)$ and $\Theta(\alpha, \beta) = 2\beta$, where $\alpha \in [0, 2\pi), \beta \in [0, 2\pi)$. Details are described in the supplementary materials.

**Conventional pulse characterization.** Figure 2 shows the conventional characterization of the intracavity-controlled OAM-carrying femtosecond laser with two common states on the HOP sphere: north pole $(0, +\pi/2)$ and the radial $(0,0)$. For the 2D spatial domain measurements on the cross-section, the experimental results and the simulations are well-matched, but not entirely. Firstly, the output modes (the total intensity distributions, Figures 2A1(vi) and 2A2(vi)) are not the standard $LG_0^{\pm 1}$ mode (Figures 2B1(vi) and 2B2(vi)) but distorted donut-like patterns; since the major axis of the intensity elliptical annulus rotates with the optical axis of the Q-plate, the intensity distributions on these annuli are not centrosymmetric about the center of the circle but rather concentrated on the two poles due to alignment imperfection. Secondly, the output polarization states are not pure. For example, in Figures 2C1(ii) and 2C1(iii), the Q and U components of the north pole state are nonzero, indicative of residual linear polarization components in the output light. Impurities within the polarization states are further visualized by the north pole and radial polarization ellipses present in Figures 2E1 and 2E2, respectively. In the north pole state (Figure 2E1), only a few perfect left-handed circular polarizations exist, and the axes of the left-handed elliptical polarization are modulated by some linear polarization components. The radial state shows cases of deteriorated linear polarization states with circular components (Figure 2E2). Therefore, the purity of both the mode and polarizations should be further improved to produce the ideal, extremely refined output pulses. Measurements for other states, including south pole, azimuthal, ($\pi/2$, $\pi/4$), (0, $-\pi/4$), ($\pi/2$, 0) and ($3\pi/2$, 0) are reported in the supplementary materials.

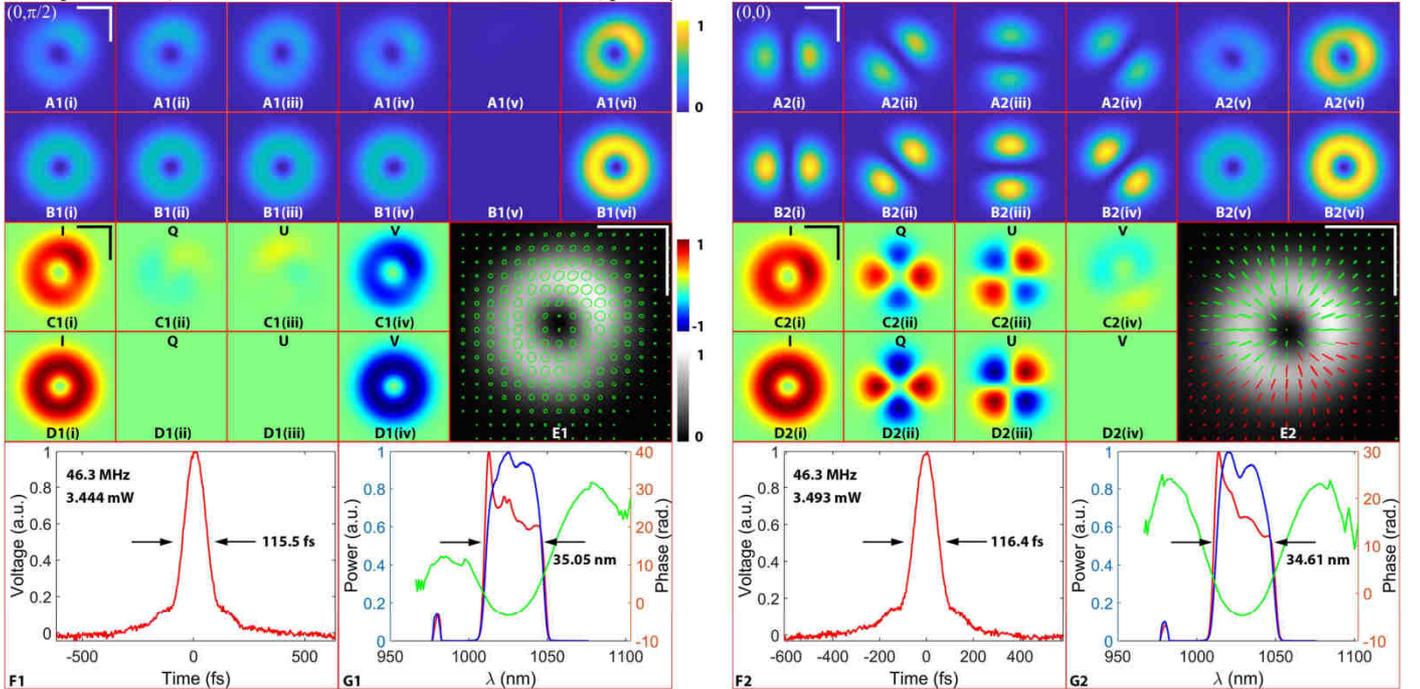

**Figure 2.** Conventional characterizations of the north pole and the radial states. **A1-G1 (left)**: $|-1, L\rangle$, the north pole $(0, \pi/2)$; **A2-G2 (right)**: $(|-1, L\rangle + |+1, R\rangle)/\sqrt{2}$, the radial state $(0,0)$; **A1(i-v)**: the 2D intensity measurements with different configurations; **A1(i-iv)**: the patterns when the polarizer with orientation at: (i) = 0°; (ii) = 45°; (iii) = 90°; (iv) = 135°; **A1(v)**: the pattern when the QWP and the polarizer are both orientated at 45°; **A1(vi)**: the total intensity distribution; **B1(i-vi)**: the corresponding simulations of A1; **C1(i-iv)**: the Stokes parameters; **D1(i-iv)**: the corresponding simulations of C1; **E1**: the polarization ellipses, where green line denotes the left-handed polarization, as the red to right-handed polarization, and blue to linear polarization; **F1** is the pulse duration measurement of the reference beam; **G1** is the spectrum of the sample beam (blue line), reference beam (red line), as well as the phase (green line). **A2-G2** follow the same description. Scale bars represent 1 mm.

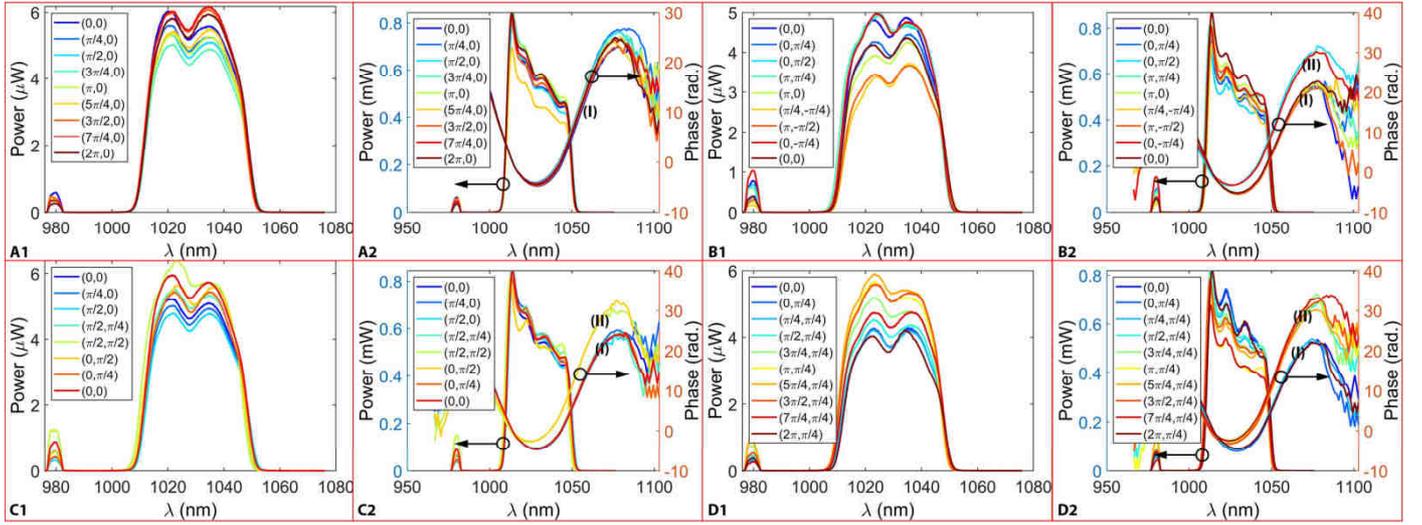

**Figure 3.** Spectrum and phase measurements of the pulse continually modulated along a closed path on the HOP sphere. **A**: the equator (Path A); **A1**: spectrum of sample pulse; **A2**: spectrum and phase of reference pulse; **B**: the circle of 0° and 180° longitude (Path B); **C**: the polar triangle between (0,0), ($\pi/2$,0) and (0,$\pi/2$) (Path C); D: the circle of 45° latitude (Path D). **B-D** follow the same description as A. Every specific spherical coordinate ($\Phi$, $\theta$) in the figure legend denotes a measurement point. I and II denote the two spectral phase zones containing the radial and the north pole state, respectively.

For the single-pixel time and spectral domain measurements, the repetition rate of the laser is ~46 MHz. The temporal duration is 110-120 fs, and the full width at half maximum (FWHM) of the spectrum is 34-36 nm. The average output power is approximately 3-5 mW. The temporal duration and spectral phase were measured from the reference beam. The spectral phase measurement is necessary for polarization-sensitive spatiotemporal pulse characterization, which will be discussed in the next section. Although not comprehensively, these measurements still reveal some of the important pulse properties especially if the output states are continuously and adiabatically modulated along the HOP sphere.

Intracavity modulation of $HOP_{SS}$ will alter the spectrum and the spectral phase due to the internal loss of the cavity. We demonstrate these in Figure 3 by continually switching the state along a certain closed path, defined as the following: Path A, the equator; Path B, the circle of 0° and 180° longitude; Path C, the polar triangle between (0,0), ($\pi/2$,0) and (0,$\pi/2$); Path D, the circle of 45° latitude, and; Path E, the circle of −45° latitude (measurements of Path E are reported in the supplementary materials). Our measurements show that the spectral phases of 45 different states uniformly distributed on the sphere fall into two zones: Zones I and II, which include the radial and the north pole state, respectively. The polarization ellipses corresponding to the 4 Paths are presented in Figure 4 to further demonstrate the pulse polarization structures as well as the capability of the femtosecond laser to generate a variety of HOP$_{SS}$ beams.

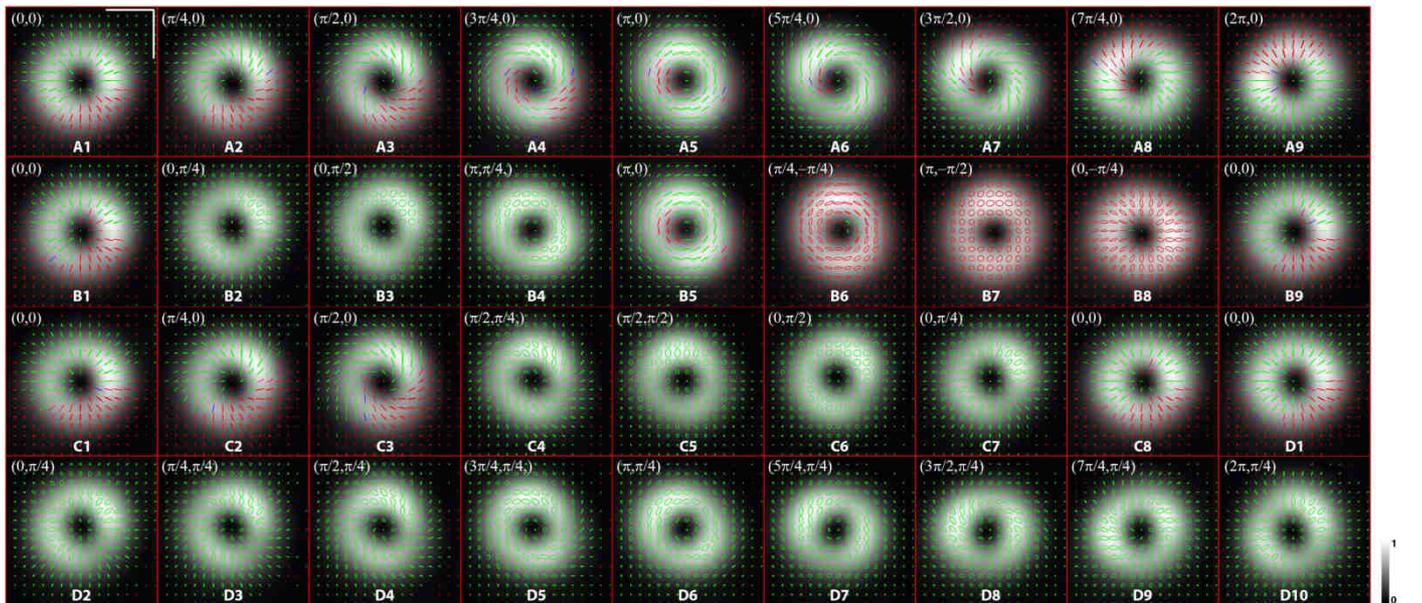

**Figure 4.** The polarization ellipses corresponding to Figure 3. **A1-9**: Path A; **B1-9**: Path B; **C1-8**: Path C; **D1-10**: Path D. Each Roman numeral represents one measurement point. All paths use the radial point state as the starting point. Scale bar represents 1 mm.

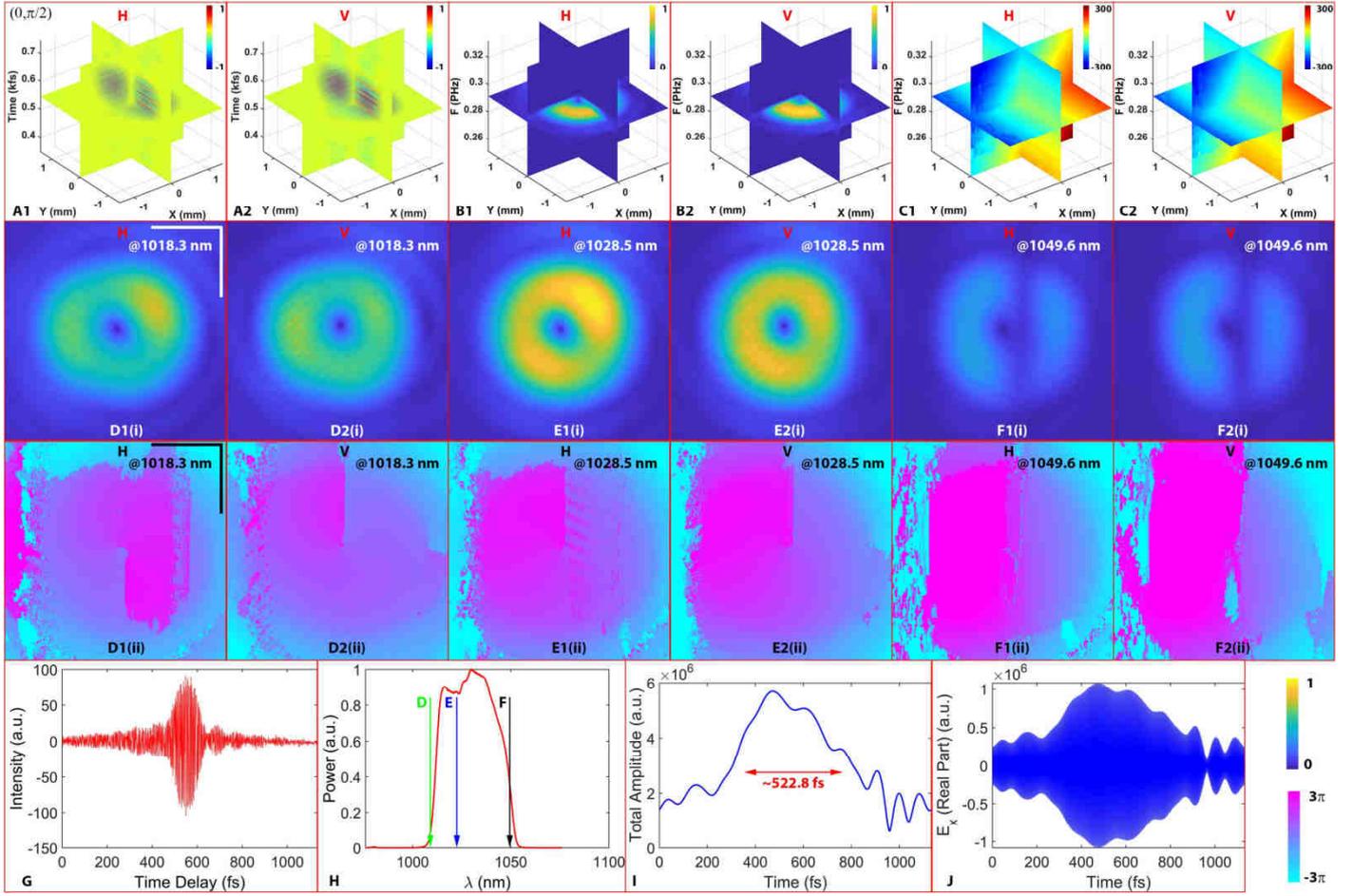

**Figure 5.** The polarization-sensitive spatiotemporal characterizations of the pulse with north pole state. **A1-2**: the 3D interferograms of the horizontal (*H*) and vertical (*V*) channels, respectively; **B1-2**: the corresponding 3D spectral amplitude of the two channels; **C1-2**: the corresponding 3D spectral phase. **D**, **E**, and **F**: the 2D amplitude (Roman numeral i) and unwrapped phase profiles (Roman numeral ii) at 1018.3 nm, 1028.5 nm and 1049.6 nm, respectively. **G**: single-pixel cross-correlation signal obtained from the time scanning procedure for *H* channel; **H**: the single-pixel spectrum; **I**: the single-pixel profile for the total temporal amplitude envelope; **J**: the corresponding real part of the pulse. Scale bars represent 1 mm.

**Characterization of Pulse Multidimensionality.** A plane wave is utilized as the reference beam as it produces a uniform interference fringe background, implying a constant spatial demodulation resolution. With uniform linear polarization distribution, it can also provide measurement of one of the two components in a 2D vector field for the sample electric field across the reference beam wavefront. When the detection plane of the CCD, denoted by Plane Π coincides with this wavefront, the projection of the sample electric field—which is essentially 2D vector field—on Plane Π can be used to characterize the states of the pulse with good approximation because the amplitude of *z'* component of the electric field (mainly governed by $\Delta\Omega_0$, the angle between the reference and the sample beam) is only approximately one percent of the *x'* or *y'* component in this experiment. As such, by providing a homogenous reference beam across the whole field of Plane Π and gaining the spectrum and spectral phase data via an optical spectrum analyzer and FROG, the complete spectral domain information of the sample pulse can be obtained by using the Equation S12 (detailed principle for polarization-sensitivity characterization described in the supplementary materials). We then can further reconstruct the temporal structure of the pulse based on the Equation S13 after an inverse Fourier transform.

Figure 5 demonstrates the polarization-sensitive spatiotemporal characterizations of the femtosecond pulse with the north pole state. Figure 5G shows a typical single-pixel cross-correlation signal obtained from the time scanning procedure. Figures 5A1-2 are the 3D interferograms recorded with the 2D time scanning for the horizontal channel (Video S1), *H*, and the vertical channel (Video S2), *V*, respectively. Due to tilted sectioning and the donut-shaped pulse feature, the 2D interferograms provide inhomogeneous intensity distribution. Figures 5B1-2 are the corresponding 3D spectral amplitudes of the two channels after applying a Fourier transform and a super-Gaussian spectral filter. Figures 5C1-2 present the 3D spectral phases, revealing the complexity of the structural phase information. The X and Y axis in Figures 5A-F coincide with *x'* and *y'*, respectively. Collectively, the measurements shown in Figure 5 illustrate both the global and local spectral structures of the complex pulse which is essentially equivalent to the corresponding temporal structures (discussed later in Figure 6). The single-pixel spectrum at different spatial points reveals microscopic differences within one channel, and even more drastic between the two channels, which is an evidence for the spatiotemporal coupling effects in the spectral domain[48] as well as the polarization structures. In addition, the dominant background of the 3D phase structure is the global phase generated by $\Delta\Omega_0$, which can be removed to uncover the local phase structure, such as the spiral wavefront. Applying the background clearing operation by subtracting out a set of specific frequency-dependent phases on Plane Π at three different frequencies (1018.3 nm, 1028.5 nm and 1049.6 nm, Figure 5H), and the resulting 2D amplitude and the corresponding unwrapped phase profiles are shown in Figures 5D-F (amplitude shown in (i), and unwrapped phase profiles shown in (ii)). The amplitude profiles from both *H* and *V* channels at all three wavelengths exhibit the donut-shaped contours that are well-matched with the 3D amplitude structures, and the unwrapped phase precisely demonstrates the spiral feature of the vortex beam carrying the helicity of −1. Figure 5I is the single-pixel signal of the temporal amplitude of the chirped pulse with the measured FWHM of

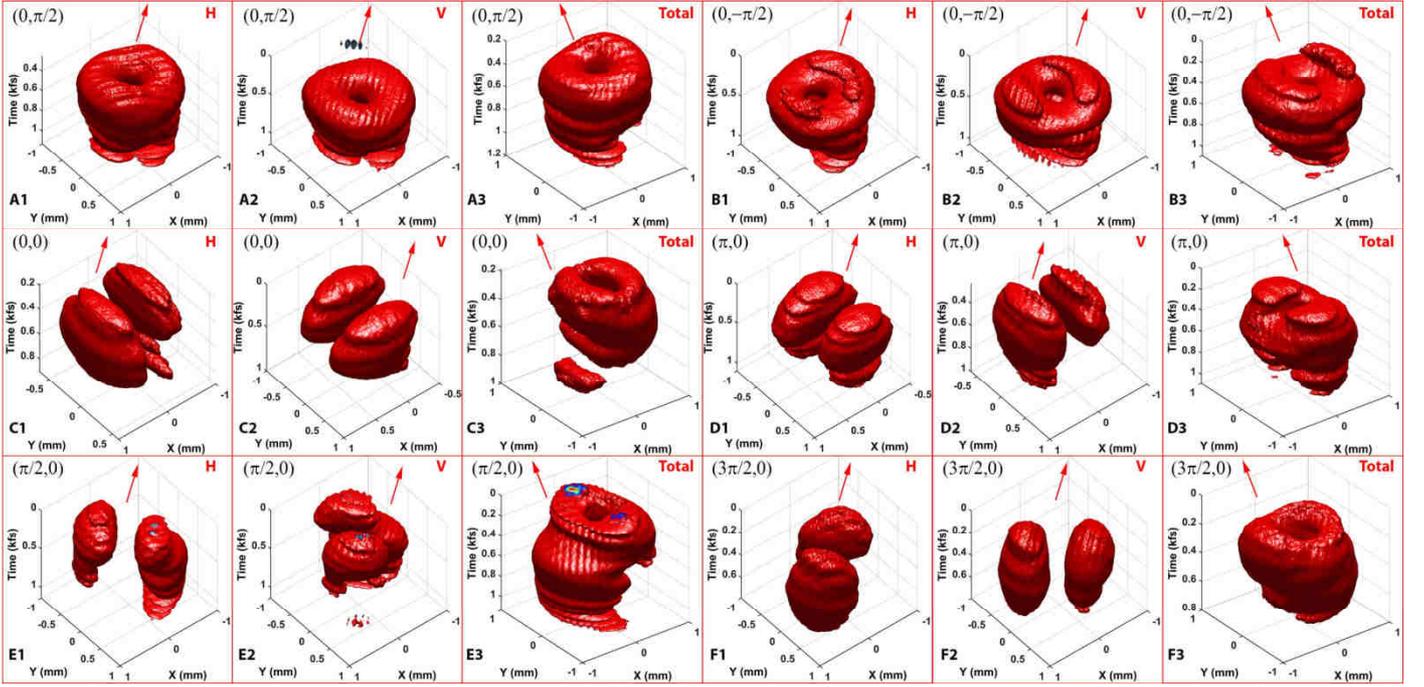

**Figure 6.** The 3D spatiotemporal reconstruction of pulse electric field $E(r,t)$ of six typical HOP$_{SS}$. **A**: the north pole state; **A1** and **A2**: amplitudes from the H and V channels, respectively; **A3**: the total amplitude; **B**: the south pole state; **C**: the radial state; **D**: the azimuthal state; **E**: the state at $(\pi/2, 0)$; **F**: the state at $(3\pi/2, 0)$. **B-F** follow the same description as A. The isosurface (red surface) shown is set at 0.4 of the maximum amplitude.

~500 fs, and Figure 5J is the corresponding real part of the H channel, which includes 8192 sampling points in order to resolve the carrying frequency.

The time scanning procedure allows for reconstruction of the pulse electric field, $E(r,t)$, in the spatiotemporal domain. The electric field (i.e., the total amplitude) is the root mean square of the two channels and is insensitive to the relative phase. Figure 6 shows the 3D reconstructions of the north pole (Videos S3-5), south pole (Videos S6-8), radial (Videos S9-11), azimuthal (Videos S12-14), $(\pi/2, 0)$ (Videos S15-17), and $(3\pi/2, 0)$ (Videos S18-20) of the HOPss. The red arrows denote the propagation direction of the pulse, $k_0$, defined by a calibration pulse with a plane wavefront, which can be realized by removing the Q-plate in the laser cavity. The temporal axis, $t$ (Time axis in Figure 6), is antiparallel with the z' axis as $\varphi = \omega_0 t - k_0 \cdot r$, where $\varphi$ is the phase and $\omega_0$ is the center angular frequency of the pulse. As shown in Figures 6A and 6B, the north pole and south pole states temporally maintain a donut-like 3D shape in both H and V channels as well as the total amplitude structure, supported by the conventional characterization method shown in Figures 2A1 and 2A2 (H channel shown in (i), V channel shown in (iii), and total amplitude shown in (vi)). This demonstrates the accumulation effects of the pulses injecting onto Plane $\Pi$. Although bearing a similar amplitude structure, minor discrepancies in both the spatial domain (the X-Y plane) and the temporal domain (the Time axis) between the two states are still noted. These discrepancies are caused by the diversity between the spectral phase shown in Figures 2G1 and 2G2. Figures 6C and 6F depict the results from additional four states along the equator, including $(0,0)$, $(\pi/2, 0)$, $(\pi, 0)$, and $(3\pi/2, 0)$. The total amplitude profiles of these states show the donut-shaped structure as anticipated. In addition, the local profile of the two sub-pulses, whose axis of symmetry rotates with respect to the state being modulated, can be visualized through the H and V channels. However, the spectral bandwidth of the output pulses is inadequate to generate a distinct twisting effect of the polarization distributions along the temporal axis—which is caused by the dispersion of the VSP—that provides different Pancharatnam-Berry phase for different wavelengths. An ultra-wideband pulse or a large geometric phase dispersive device can be utilized to visualize the twisting effect, such as a pulse temporally transformed from radial to azimuthal polarization.

In our multidimensional analysis of pulse, all amplitude profiles affirm a donut-shaped pulse that is supported by the 2D spatial structure prediction. Nevertheless, the intrinsic spectral phase will decidedly modify the temporal structure, as shown in Figures 6A and 6C. The polarization resolvability of the system is further verified by the different spatiotemporal structures from the H and V channels within the measured HOPss pulse. For instance, Figures 6C-F demonstrate that specific spatial features of a pulse are determined by its vectorial properties, and that the temporally accumulated pattern is consistent with the 2D conventional measurement. It is worth noting that the system can resolve any tilted input sample light fields as long as the angle between the signal propagation direction and the normal of Plane $\Pi$ is within ±3.18°. Detailed descriptions are reported in the supplementary materials.

## Discussion

We have presented a femtosecond pulse scope that unites vector vortex mode-locked lasing and vectorial quantification. By controlling the Pancharatnam-Berry phase via the VSP and QWP within the laser cavity, OAM-carrying beams (i.e., HOPss) can be produced and continually modulated while imposing minimal effect on the status of the passive mode-locked operation. Most of the laser configurations described above are self-start. However, it is generally more favorable to achieve a given state through self-starting at the radial state $(0,0)$ then continuously modulating to the target state. Given a diode-pumped light with adequate power, the mode-locked operation remains stable for weeks. The presence of the free-space section does not degrade the long-term stability of the laser. In practice, the intracavity configuration can be conveniently adapted into existing mode-locked lasers, enabling a flexible tailoring of the structured light in spatial, temporal, and spectral domains, as well as polarization distribution. With these unique features, this pulse-scope system can provide a unique light source for spatiotemporal nonlinear optics, laser machining, nanoparticle manipulation, and spatial mode division multiplexed systems.

Laser pulses in either laboratory or industry are typically complex objects. Unlike the classical electromagnetic (EM) wave with uniform polarization distribution along a flat wavefront or a general vector beam under the paraxial approximation, real-world light pulses, such as beams from a high-power multimode fiber laser, dechirped femtosecond pulses with structured features, or even more intricate optical phenomena (e.g., knot-shaped pulses), usually carry a non-vanishing component in the propagation direction. Therefore, the description of a general vectorial laser pulse should ideally be implemented in a multidimensional way. Herein, we have introduced the initial concept and implementation of a polarization-sensitive temporal scanning interferometer to capture the sophisticated information of these complex HOP$_{SS}$ pulses. The pulse-scope system provides the information on spatiotemporal amplitude and phase distribution as well as the polarization-sensitive features of the complex laser pulses.

Like most laser systems, mechanical instability and environmental vibrations may degrade the measurement accuracy. The pulse scope characterizes the beam without the relative phase between the $H$ and $V$ channels; the capability of recording null relative phase information can further enable spatiotemporal reconstruction of a complex pulse as a 3D vector field. The quantification ability of this pulse-scope system, nevertheless, assures a new means of exploration for in-depth characterization and optimization of ultrafast laser pulse, paving the way for investigating novel pulse generations as well as for studying the interactions between materials and structured pulses. This integrated device will facilitate adaptive customization of structured pulses that can be applied in optical physics research and laser-based manufacturing.

## Material and methods

**Intracavity-controlled OAM-carrying femtosecond laser.** The laser system (gray portion of the floor, Figure 1) has a unidirectional ring cavity employed for self-starting operation, in which the propagation direction is denoted by the red circular arrow. The gain medium, a 26-cm long Ytterbium-doped fiber, is pumped by a low-profile laser diode which can deliver up to 500 mW (at 980 nm) fiber-coupled kink free output power through a 980/1030-nm wavelength-division multiplexing coupler. Passing through QWP-I, the polarizing beamsplitter (PBS, extinction ratio of 1000:1), QWP-III, and the VSP, the light launched from the transmitting collimator (collimator-Tx) is reflected back to the cavity by the output coupler with reflectivity of 90% and the unprotected gold mirror II (Mirror-II). Two pairs of reflective diffraction gratings (600 lines/mm, 1-μm blaze) are used to compensate for the normal group-velocity dispersion of the fiber. The PBS—which maintains its p-polarization along $x$—QWP-III, the VSP, the output coupler, Mirror-II, and half wave plate I (HWP-I) composes a HOP sphere beams generator that produce the on-demand HOP$_{SS}$ from the pure, horizontal ($x$) linear polarization states. From the output coupler, the reflected light then transforms its polarization distribution back to the linear polarization state which can be expressed as $-\cos(2\beta)\hat{e}_x + \sin(2\beta)\hat{e}_y$; that is, the polarization direction has a degree of 2β over $x$, assuming the angle between the input and reflected light (~0.85°) is negligible (detailed in the supplementary materials). Here, HWP-II is employed to rotate the polarization of the reflected light from 2β to 0 to further reduce the additional loss of the whole cavity because it is crucial to ensure that the vibrational direction of the light launched to the reflected diffraction grating is perpendicular to the direction of the groove due to the polarization-sensitive property of the grating. . The PBS also provides the reference plane wave which carries the vertical ($y$) linear polarization for the polarization-sensitive spatiotemporal pulse characterization system. The light then passes through the isolator, HWP-I, QWP-I before collected by the receiving collimator (collimator-Rx) into the 130-cm single-mode fiber (SMF). Mode-locked operation is stabilized by nonlinear polarization evolution which is implemented with the bulk wave plates (HWP-I, QWP-I and QWP-II) and SMF.

**Conventional pulse characterization.** To identify the pulse duration of the reference beam (pink arrow, Figure 1), we assumed the reference beam is transversely homogeneous and thus has a well-defined pulse duration relatively to the complex sample pulse. Thus, a custom-built interferometric autocorrelator equipped with a two-photon photodetector is used to measure the reference beam pulse duration.

The conventional measurement of the polarization distribution was realized through a 2D imaging polarimeter, which consists of a complementary metal-oxide-semiconductor (CMOS) camera, a polarizer, and a QWP. The reference frame is calibrated by a single-point, free-space polarimeter system. The measurement procedure is standard, which is detailed in the supplementary materials. The spatial distribution of the Stokes parameters or polarization ellipses only describes the average vectorial properties of the pulse on the cross section, and cannot provide the temporal information. From the spectral domain perspective, these measured Stoke parameters are the weighted summation of all common Stokes parameters corresponding to a single wavelength.

**Characterization of Pulse Multidimensionality.** In the multidimensional analysis, a 2D polarization-sensitive temporal scanning Mach–Zehnder interferometer (green portion of the floor, Figure 1) was used to capture the complete information of the pulses. The reference beam, exported from the PBS, is first time-delayed via an unprotected gold right-angle prism mirror, an unprotected gold hollow roof prism mirror, and a linear stage, then attenuated to a suitable intensity by a round variable reflective neutral density filter (NDF-I). Thereafter, the beam passes through a flat optical window (OW-I) for precisely balancing the dispersion between the two arms, and the polarizer rotates automatically by a motorized rotation stage. The reference beam is then expended by a laboratory-made 10x beam expender consisting of Lens-I (f = 7.5 mm), Lens-II (f = 75 mm), and a 5-μm pinhole, which selects a small portion of the energy of the reference beam around the focus, before reaching the beamsplitter (BS). The sample beam is launched from the output coupler, and then propagates through NDF-II, OW-II, Mirror-IV and -V, and finally interferes with the reference beam at the BS. Mirror-IV and -V are used to compensate for the extra optical pathlength due to the delay line in the reference arm. A CCD with a sensor size of 1034×768 pixels and a unit cell size of 4.65×4.65 μm sequentially captures the 2D interference patterns. Temporal scanning, containing 1,700 steps, is achieved by discretely translating the hollow roof prism on the linear stage. The scanning step size of 100 nm corresponds to the optical path delay ($\delta z$) of 200 nm due to the double path configuration (the right-angle prism and the hollow roof prism), equivalent to a spectral detection range ($\Delta \lambda$) of 5.3 μm. The spectral frequency resolution ($\delta \lambda$) is 3.1 nm calculated using the whole delay range ($\Delta z$) of 340 μm, and the spatial resolution of the system ($\delta x$) of 37.2 μm accords with $\Delta \Omega_0$ of ~1.6°. The maximum resolvable angle ($\Delta \Omega_{max}$) is 6.36°, determined by the pixel size of the CCD and Nyquist–Shannon sampling theorem. Generally, the EM fields can be represented by a superposition of plane waves; that is, the complex pulse is the combination of plane waves carrying different propagation directions (wave vector $\boldsymbol{k}$) but with small differences under the paraxial approximation. Therefore, this system meets the requirement for resolving the sample complex pulses and even scattered light, in which the small scattering angle carrying forward scattered light are the dominant portion.

## Acknowledgements

This work was supported by the National Institutes of Health (R01HL-125084, R01HL-127271, R01EY-026091, R01EY-028662, and P41EB-01890) and the Air Force Office of Scientific Research (FA9550-17-1-0193). C.J. was supported by the National Science Foundation (DGE-1839285); L.Y. was supported by the American Heart Association (18PRE34050021). In addition, the authors would like to thank Jing-Gao Zheng, PhD and Buyun Zhang, PhD for their meaningful discussion.

## Conflict of interest

The authors declare no competing financial interests

## Contributions

T.H., L.Q. and Z.C. conceived and designed the research. Z.C. supervised the project. T.H. performed the experiments, data processing, numerical models, and numerical analysis. J.C. and Y.M. performed the simulation. L.Q., J.C., Y.M., Y.L., and Z.Z. assisted in experiments. All the authors discussed the results and prepared the manuscript.

# References

1. Berestetskii, V. B., Pitaevskii, L. P., Lifshitz, E.M. Quantum Electrodynamics, 2$^{nd}$ edn, Vol. 4 (eds V. B. Berestetskii, L. P. Pitaevskii, E.M. Lifshitz) Ch. 1 (Butterworth-Heinemann, 1982).
2. Kogelnik H. & Li, T. Laser Beams and Resonators. *Appl. Opt.* **5**, 1550-1567 (1966).
3. Allen, L., Beijersbergen, M. W., Spreeuw, R. J. C., & Woerdman, J. P. Orbital angular momentum of light and the transformation of Laguerre-Gaussian laser modes. *Phys. Rev. A* **45**, 8185 (1992).
4. Barnett S. M. & Allen, L. Orbital angular momentum and nonparaxial light beams. *Optics Communication* **110**(5-6), 670-8 (1994).
5. Milione, G, Sztul, H. I., Nolan, D. A. & Alfano, R. R. Higher-Order Poincaré Sphere, Stokes Parameters, and the Angular Momentum of Light. *Phys. Rev. Lett.* **107**(5) 053601 (2011).
6. Simpson, N. B., Dholakia, K., Allen, L. & Padgett, M. J. Mechanical equivalence of spin and orbital angular momentum of light: An optical spanner. *Opt. Lett.* **22**(4), 52-54 (1997).
7. Wang, J. *et al.* Terabit free-space data transmission employing orbital angular momentum multiplexing. Nat. Photonics 6, 488 (2012).
8. Wong L. J. & Kärtner, F. X. Direct acceleration of an electron in infinite vacuum by a pulsed radially-polarized laser beam. *Opt. Express* **18**, 25035 (2010).
9. Török, P. & Munro, P.R.T. The use of Gauss-Laguerre vector beams in STED microscopy. *Opt. Express* **12**, 3605-3617 (2004).
10. Hao, X., Kuang, C., Wang, T. & Liu, X. Effects of polarization on the deexcitation dark focal spot in STED microscopy. *J. Opt.* **12**, 115707 (2010).
11. Friese, M., Enger, J., Rubinsztein-Dunlop, H. & Heckenberg, N. Optical angular-momentum transfer to trapped absorbing particles. *Phys. Rev. A* **54**, 1593–1596 (1996).
12. Friese, M., Nieminen, T., Heckenberg, N. R. & Rubinsztein-Dunlop, H. Optical alignment and spinning of laser-trapped microscopic particles. *Nature* **394**, 348–350 (1998).
13. Hamazaki, J. et al. Optical-vortex laser ablation. *Opt. Express* **18**, 2144–2151 (2010).
14. Lavery, M. P. J., Speirits, F. C., Barnett, S. M. & Padgett, M. J. Detection of a spinning object using light's orbital angular momentum. *Science* **341**, 6145 (2013).
15. Uribe-Patarroyo, N., Fraine, A., Simon, D. S., Minaeva, O. & Sergienko, A. V. Object identification using correlated orbital angular momentum states. *Phys. Rev. Lett.* **110**, 043601 (2013).
16. A. Mair, A. Vaziri, G. Weihs &A. Zeilinger, Entanglement of the orbital angular momentum states of photons, *Nature* **412**, 313 (2001).
17. Ndagano, B. *et al.* Characterizing quantum channels with non-separable states of classical light, *Nat. Phys.***13**,397–402 (2017).
18. Milione, G. *et al.* Cylindrical vector beam generation from a multi elliptical core optical fiber. *Proc. SPIE* **7950**, 79500K (2011).
19. Milione, G., Evans, S. E., Nolan, D. A. & Alfano, R. R. Higher Order Pancharatnam-Berry Phase and the Angular Momentum of Light. *Phys. Rev. Lett.* **108**, 190401 (2012).
20. Marrucci, L., Manzo, C., & Paparo, D. Optical Spin-to-Orbital Angular Momentum Conversion in Inhomogeneous Anisotropic Media. *Phys. Rev. Lett.* **96**, 163905 (2006).
21. Cardano, F., Karimi, E., Slussarenko, S., Marrucci, L., de Lisio, C. & Santamato, E. Polarization pattern of vector vortex beams generated by q-plates with different topological charges. *Appl. Opt.* **51**, C1-C6 (2012).
22. Devlin, R. C., Ambrosio, A., Rubin, N. A., Mueller, J. P. B. & Capasso, F. Arbitrary spin-to–orbital angular momentum conversion of light. *Science* **358**, 896 (2017).
23. W. J. Lai, B. C. Lim, P. B. Phua, K. S. Tiaw, H. H. Teo, and M. H. Hong, Generation of radially polarized beam with a segmented spiral varying retarder. *Opt. Express* **16**(20), 15694 (2008).
24. Bisson, J. F., Li, J., Ueda, K. & Senatsky, Y. Generation of Laguerre–Gaussian modes in Nd:YAG laser using diffractive optical pumping Laser. *Phys. Lett.* **2** 327–33 (2005).
25. Khonina, S. N., & Karpeev, S. V. Grating-based optical scheme for the universal generation of inhomogeneously polarized laser beams. *Appl. Opt.* **49**, 1734-1738 (2010).
26. Kim, J. W., Mackenzie, J. I., Hayes, J. R. & Clarkson, W. A. High power Er:YAG laser with radially-polarized Laguerre– Gaussian (LG01) mode output. *Opt. Express* **19** 14526–31 (2011).
27. Ito, A., Kozawa, Y. & Sato, S. Generation of hollow scalar and vector beams using a spot-defect mirror. *J. Opt. Soc. Am. A* **27** 2072–7 (2010).
28. Ngcobo, S., Litvin, I., Burger, L. & Forbes, A. A digital laser for on-demand laser modes. *Nat. Commun.* **4** 2289 (2013).
29. Naidoo, D. *et al.* Controlled generation of higher-order Poincaré sphere beams from a laser. *Nat. Photonics* **10**, 327 (2016).
30. Zhang, Y. *et al.* Self-mode-locked Laguerre-Gaussian beam with staged topological charge by thermal-optical field coupling. *Opt. Express* **24**, 5514-5522 (2016).
31. Mao D. *et al.* Ultrafast all-fiber based cylindrical-vector beam laser. *Appl. Phys. Lett.* **110**, 021107 (2017).
32. Huang, K. Zeng, J. Gan, J. Hao, Q. & Zeng, H. Controlled generation of ultrafast vector vortex beams from a mode-locked fiber laser. *Opt. Lett.* **43**(16), 3933–3936 (2018).
33. Zhao Y. *et al.* Intracavity cylindrical vector beam generation from all-PM Er-doped mode-locked fiber laser. *Opt. Express* **27**, 8808-8818 (2019).
34. Wang S *et al.* Direct emission of chirality controllable femtosecond LG01 vortex beam. *Appl. Phys. Lett.* **112**, 201110 (2018).
35. Wang, T. *et al.* Generation of Femtosecond Optical Vortex Beams in All-Fiber Mode-Locked Fiber Laser Using Mode Selective Coupler. *Lightwave Technol.* **35**, 2161 (2017).
36. Hamamatsu: FESCA (FESCA-100 Femtosecond streak camera: C11853-01, information sheet, August 2019) and actual information under: https://www.hamamatsu.com/us/en/product/photometry-systems/streak-camera/fesca-100-femtosecond-streak-camera/index.html
37. Weber, H. P. Method for pulsewidth measurement of ultrashort light pulses, using nonlinear optics. *J. Appl. Phys.* **38**, 2231 (1967).
38. Trebino, R. & Kane, D. J. Using phase retrieval to measure the intensity and phase of ultrashort pulses: frequency-resolved optical gating. *J. Opt. Soc. Am. A* **10**, 1101-1111 (1993).
39. Iaconis, C. & Walmsley, I. A. Spectral phase interferometry for direct electric-field reconstruction of ultrashort optical pulses. *Opt. Lett.* **23**, 792-794 (1998).
40. Lozovoy, V. V., Pastirk, I. & Dantus, M. Multiphoton intrapulse interference. IV. Ultrashort laser pulse spectral phase characterization and compensation. Opt. Lett. 29, 775-777 (2004).
41. Miranda, M., Fordell, T., Arnold, C., L'Huillier, A. & Crespo H. Simultaneous compression and characterization of ultrashort laser pulses using chirped mirrors and glass wedges. *Opt. Express* **20**, 688-697 (2012).
42. Akturk, S., Kimmel, M., O'Shea, P. & Trebino, R. Measuring spatial chirp in ultrashort pulses using single-shot Frequency-Resolved Optical Gating. *Opt. Express* **11**, 68-78 (2003).
43. Gallmann, L. *et al.* Spatially resolved amplitude and phase characterization of femtosecond optical pulses. *Opt. Lett.* **26**, 96-98 (2001).
44. Bowlan, P., Gabolde, P. & Trebino, R. Directly measuring the spatio-temporal electric field of focusing ultrashort pulses. *Opt. Express* **15**, 10219-10230 (2007).
45. Gabolde, P. & Trebino, R. Self-referenced measurement of the complete electric field of ultrashort pulses. *Opt. Express* **12**, 4423-4429 (2004).
46. Grunwald, R., Neumann, U., Griebner, U., Reimann, K., Steinmeyer, G. & Kebbel, V. Ultrashort-pulse wave-front autocorrelation. *Opt. Lett.* **28**, 2399-2401 (2003).
47. Miranda, M. *et al.* Spatiotemporal characterization of ultrashort laser pulses using spatially resolved Fourier transform spectrometry. *Opt. Lett.* **39**, 5142-5145 (2014).
48. Pariente, G., Gallet, V., Borot, A., Gobert, O. & Quéré, F. Space–time characterization of ultra-intense femtosecond laser beams. *Nat. Photon.* **10**, 547–553 (2016).
49. Suzuki, M., Yamane, K., Oka, K., Toda, Y. & Morita, R. Extended Stokes parameters for cylindrically polarized beams. *Optical Review* **22**, 179–183 (2015).
50. Mitchell, K. J., Radwell, N., Franke-Arnold, S., Padgett, M. J. & Phillips, D. B. Polarisation structuring of broadband light. *Opt. Express* **25**, 25079-25089 (2017).

# Supplementary Material: Integrated pulse scope for tunable generation and intrinsic characterization of structured femtosecond laser

Tiancheng Huo, Li Qi, Jason J. Chen, Yusi Miao, Yan Li, Zhikai Zhu & Zhongping Chen

Beckman Laser Institute and Medical Clinic, Department of Biomedical Engineering, University of California, Irvine, Irvine, CA 92612, USA

In this supplementary information, we provide detailed descriptions of this integrated system for intracavity generation and multidimensional characterization of broadband HOP$_{SS}$ femtosecond pulses, and present validation tests of the measurement results. This document is organized as follows:

In Section I, we describe the calculations for the laser cavity set-up and the method for the Stokes parameters measurements as well as its data processing.

In Section II, we detail the numerical processing technique applied to the raw experimental data to obtain the spatiospectral and spatiotemporal properties of the femtosecond pulses.

## Section I: Calculations for the laser cavity setup and method for the Stokes parameters measurements as well as its data processing

### I.1. Calculating the map between the HOP$_{SS}$ and the parameter $\alpha$, $\beta$

Geometric phase-based elements, such as Q-plates with different topological charge, J-plates, and sectored spatially varying retarders, can provide a direct connection between the SAM and OAM which serves as a spin-orbit-converter. This conversion depends on the symmetricity of the local optical axis distribution on the device, and it transforms a well-defined monochromatic plane electromagnetic (EM) wave with pure polarization states descripted by arbitrary points on a common Poincaré sphere to the states on the HOP sphere, where the two poles are generated by the two monochromatic plane EM wave with pure circular polarization states carrying opposite helicity (left and right circular polarization states: |L⟩ and |R⟩). Here, the laser is equipped with Q-plate (q = +1/2), the most common geometric phase device, to perform the transformation where the left and right circular polarizations are converted to output states with opposite spin and $\pm 2q\hbar$ OAM, as follows:

$$|L\rangle \to |+2q, R\rangle = |+2q\rangle \otimes |R\rangle = \exp(+i2q\varphi)|R\rangle$$
$$|R\rangle \to |-2q, L\rangle = |-2q\rangle \otimes |L\rangle = \exp(-i2q\varphi)|L\rangle \quad (S1)$$

where $|\pm 2q\rangle = \exp(\pm i2q\varphi)$ is the azimuthal coordinate, and $\varphi = \arctan(y/x)$.

By modulating the angles between the x-axis of the laboratory system ($x$) and the optical axis of QWP-III, $\alpha$, and between $x$ and that of the Q-plate, $\beta$, the input horizontal polarization states are mapped onto the entire mean HOP sphere $[\bar{S}^{+1}(\Phi(\alpha,\beta), \Theta(\alpha,\beta))]$, where $\Phi$ and $\Theta$ are the polar and azimuthal angles of the spherical coordinates, respectively.

Using the Jones calculus, the input state after the PBS can be the expressed as:

$$|\text{Input}\rangle = E_{x0}(t)e^{i[\varphi_{x0}(t)-\omega_0 t]}\hat{e}_x = E_{x0}(t)e^{i[\varphi_{x0}(t)-\omega_0 t]}\begin{bmatrix}1\\0\end{bmatrix}$$

where $\omega_0$ is the central angular frequency, and $\varphi_{x0}(t)$ and $E_{x0}(t)$ are the phase and amplitude corresponding to the envelope, respectively. $\hat{e}_x$ is the unit vector of $x$-axis of the laboratory system ($oxyz$). Furthermore, the Q-plate and the QWP have the matrix formulation as:

$$M_{Q-plate} = \begin{bmatrix}\cos(2q\varphi+2\alpha) & \sin(2q\varphi+2\alpha)\\ \sin(2q\varphi+2\alpha) & -\cos(2q\varphi+2\alpha)\end{bmatrix}$$

$$M_{QWP} = \begin{bmatrix}\cos^2(\beta)+i\sin^2(\beta) & \sqrt{2}e^{-i\frac{\pi}{4}}\sin(\beta)\cos(\beta)\\ \sqrt{2}e^{-i\frac{\pi}{4}}\sin(\beta)\cos(\beta) & \sin^2(\beta)+i\cos^2(\beta)\end{bmatrix}$$

Therefore, the output states $|\bar{S}^{+1}(\Phi(\alpha,\beta),\Theta(\alpha,\beta))\rangle$ can be calculated as:

$$|\bar{S}^{+1}(\Phi(\alpha,\beta),\Theta(\alpha,\beta))\rangle = M_{Q-plate}M_{QWP}|\text{Input}\rangle = \frac{1}{\sqrt{2}}E_{x0}(t)e^{i[\varphi_{x0}(t)-\omega_0 t]}\begin{bmatrix}e^{-i\frac{\pi}{4}}\cos(2q\varphi-2\beta+2\alpha)+e^{+i\frac{\pi}{4}}\cos(2q\varphi+2\alpha)\\ e^{-i\frac{\pi}{4}}\sin(2q\varphi-2\beta+2\alpha)+e^{+i\frac{\pi}{4}}\sin(2q\varphi+2\alpha)\end{bmatrix}$$

In the picture of the circular polarization, this equation can be expressed as:

$$|\bar{S}^{+1}(\Phi(\alpha,\beta),\Theta(\alpha,\beta))\rangle = \sin(\frac{\Theta}{2}+\frac{\pi}{4})e^{-i\frac{\Phi}{2}}|-1,L\rangle + \cos(\frac{\Theta}{2}+\frac{\pi}{4})e^{+i\frac{\Phi}{2}}|+1,R\rangle \quad (S2)$$

where |L⟩ and |R⟩ have the following forms in the picture of the linear polarization:

$$|L\rangle = \frac{1}{\sqrt{2}}E_{x0}(t)e^{i[\varphi_{x0}(t)-\omega_0 t]}\begin{bmatrix}1\\i\end{bmatrix} \text{ and } |R\rangle = \frac{1}{\sqrt{2}}E_{x0}(t)e^{i[\varphi_{x0}(t)-\omega_0 t]}\begin{bmatrix}1\\-i\end{bmatrix}$$

The relationship between $\Phi$, $\Theta$ and $\alpha$, $\beta$ is:

$$\Phi(\alpha,\beta) = 2(\alpha-\beta),$$
$$\text{where } \Theta(\alpha,\beta) = 2\beta, \text{ and } \alpha \in [0,2\pi), \beta \in [0,2\pi) \quad (S3)$$

## I.2. Calculation for the light reflected state from the output coupler

With respect to reflected direction (reverse of the $oz$ direction) of the femtosecond pulse, the Q-plate, the QWP and the output coupler have the matrix formulations shown as follow:

$$\bar{M}_{Q-plate} = \begin{bmatrix} \cos(2q\varphi + 2\alpha) & -\sin(2q\varphi + 2\alpha) \\ -\sin(2q\varphi + 2\alpha) & -\cos(2q\varphi + 2\alpha) \end{bmatrix}$$

$$\bar{M}_{QWP} = \begin{bmatrix} \cos^2(\beta) + i\sin^2(\beta) & -\sqrt{2}e^{-i\frac{\pi}{4}}\sin(\beta)\cos(\beta) \\ -\sqrt{2}e^{-i\frac{\pi}{4}}\sin(\beta)\cos(\beta) & \sin^2(\beta) + i\cos^2(\beta) \end{bmatrix}$$

$$\bar{M}_{output-coupler} = \begin{bmatrix} -1 & 0 \\ 0 & 1 \end{bmatrix}$$

Therefore, the back-reflected signal can be calculated as:

$$|\text{Reflected}\rangle = \bar{M}_{QWP}\bar{M}_{Q-plate}\bar{M}_{output-coupler}|\bar{S}^{+1}(\Phi(\alpha,\beta),\Theta(\alpha,\beta))\rangle = \begin{bmatrix} -\cos(2\beta) \\ \sin(2\beta) \end{bmatrix} \quad \text{(S4)}$$

which means the polarization direction has a degree of double $\beta$ over $x$ if one ignores the small angle between the input and reflected light (~0.85°).

## I.3. Principles for the Stokes parameters measurements

If the monochromatic electric field is described as:

$$\mathbf{E}(\mathbf{r},t) = \begin{bmatrix} E_x(\mathbf{r},t) \\ E_y(\mathbf{r},t) \end{bmatrix} = \begin{bmatrix} E_{x0}e^{-i(\omega t - \mathbf{k}\cdot\mathbf{r} + \varphi_x)} \\ E_{y0}e^{-i(\omega t - \mathbf{k}\cdot\mathbf{r} + \varphi_y)} \end{bmatrix}$$

where $\omega$ is the angular frequency, $\mathbf{k}$ is the wave vector, $E_{x0}$ and $E_{y0}$ are the constant amplitude, and $\varphi_x$ and $\varphi_y$ are the constant phase, then the definition of the Stokes parameters $(I, Q, U, V)$ is:

$$\begin{cases} I = E_{y0}^2 + E_{x0}^2 \\ Q = E_{y0}^2 - E_{x0}^2 \\ U = 2E_{y0}E_{x0}\cos(\varphi_{yx}) \\ V = 2E_{y0}E_{x0}\sin(\varphi_{yx}) \end{cases} \quad \text{(S5)}$$

where $\varphi_{yx} = \varphi_y - \varphi_x$.

The characterization system for the Stokes parameters includes three parts: the QWP, the linear polarizer, and a CCD, as shown in Figure S1. The measurement procedures are shown in the Table S1. First, a linear polarizer is placed in front of the CCD with its fast axis at 0° to record $E_{x0}^2$, then at 45° to record $\frac{1}{2}[E_{y0}^2 + E_{x0}^2 + 2E_{y0}E_{x0}\cos(\varphi_{yx})]$, at 90° to record the $E_{y0}^2$. This is followed by placing a QWP with its fast axis at 45° and the same linear polarizer at 45° in front of the CCD to obtain the signal $\frac{1}{2}[E_{y0}^2 + E_{x0}^2 - 2E_{y0}E_{x0}\sin(\varphi_{yx})]$.

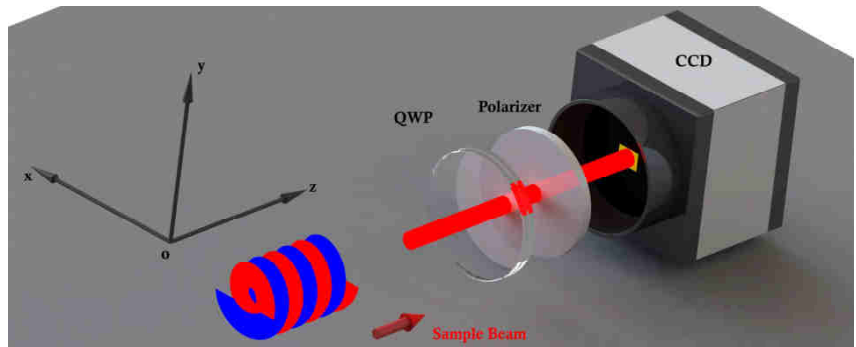

**Figure S1.** Schematic diagram of the characterization system for the Stokes parameters. QWP: quarter-wave plate. The reference systems $oxyz$ denotes the laboratory system, and the positive direction of $oz$ refers to the propagation direction of the input pulses (denoted by the red arrow).

| Step | Configuration | Measured Signal | Relationship between the measured signal and its amplitude and the phase |
|------|---------------|-----------------|--------------------------------------------------------------------------|
| 1 | Polarizer @ 0° | $I_{0°}$ | $I_{0°} = E_{x0}^2$ |
| 2 | Polarizer @ 45° | $I_{45°}$ | $I_{45°} = \frac{1}{2}[E_{y0}^2 + E_{x0}^2 + 2E_{y0}E_{x0}\cos(\varphi_{yx})]$ |

| 3 | Polarizer @ 90° | $I_{90°}$ | $I_{90°} = E_{y0}^2$ |
| 4 | QWP @ 45° (before) + Polarizer @ 45° (after) | $I_{QP}$ | $I_{QP} = \frac{1}{2}[E_{y0}^2 + E_{x0}^2 - 2E_{y0}E_{x0}\sin(\varphi_{yx})]$ |

**Table S1.** The procedures for the Stokes parameters measurements.

Based on the table 2, the expression for the Stokes parameters can be obtain:

$$\begin{cases} I = I_{90°} + I_{0°} \\ Q = I_{90°} - I_{0°} \\ U = 2I_{45°} - (I_{90°} + I_{0°}) \\ V = (I_{90°} + I_{0°}) - 2I_{QP} \end{cases} \quad (S6)$$

Because the femtosecond pules are broadband signal, the Stokes parameters demonstrate the accumulated effects that the results are the summation of the different Stokes parameter corresponding to the different single wavelength.

### I.3. Conventional characterization of the intracavity-controlled OAM-carrying femtosecond laser

In this section, more experimental results for the conventional characterization are presented. All results are grouped into five parts corresponding to the five modulation paths along the HOP sphere, which is shown in the Figure S2. The five paths include: Path A, the equator or the red circle; Path B, the circle of 0° and 180° longitude, or the green circle; Path C, the polar triangle between (0,0), (π/2,0) and (0,π/2); Path D, the circle of 45° latitude, or the blue circle with $\Theta = \pi/4$, and; Path E, the circle of −45° latitude, or the cyan circle with $\Theta = -\pi/4$. $\alpha$ and $\beta$ each has an offset of −8° and −4.9°, respectively.

### I.3.A. Path A, the Equator

This part demonstrates the detailed characterizations for the states modulated along the path A. Table S2A shows the theoretical (denoted by the letter $T$) and experimental (denoted by the letter $E$) values of the parameters $\Phi, \Theta, \alpha, \beta$, the output power, as well as the coefficients ($c_1$ and $c_2$) for two eigenstates, $|-1, L\rangle$ and $|+1, R\rangle$. Figures S3A.1-3 are the conventional characterizations of the intracavity controlled OAM-carrying femtosecond laser pulses.

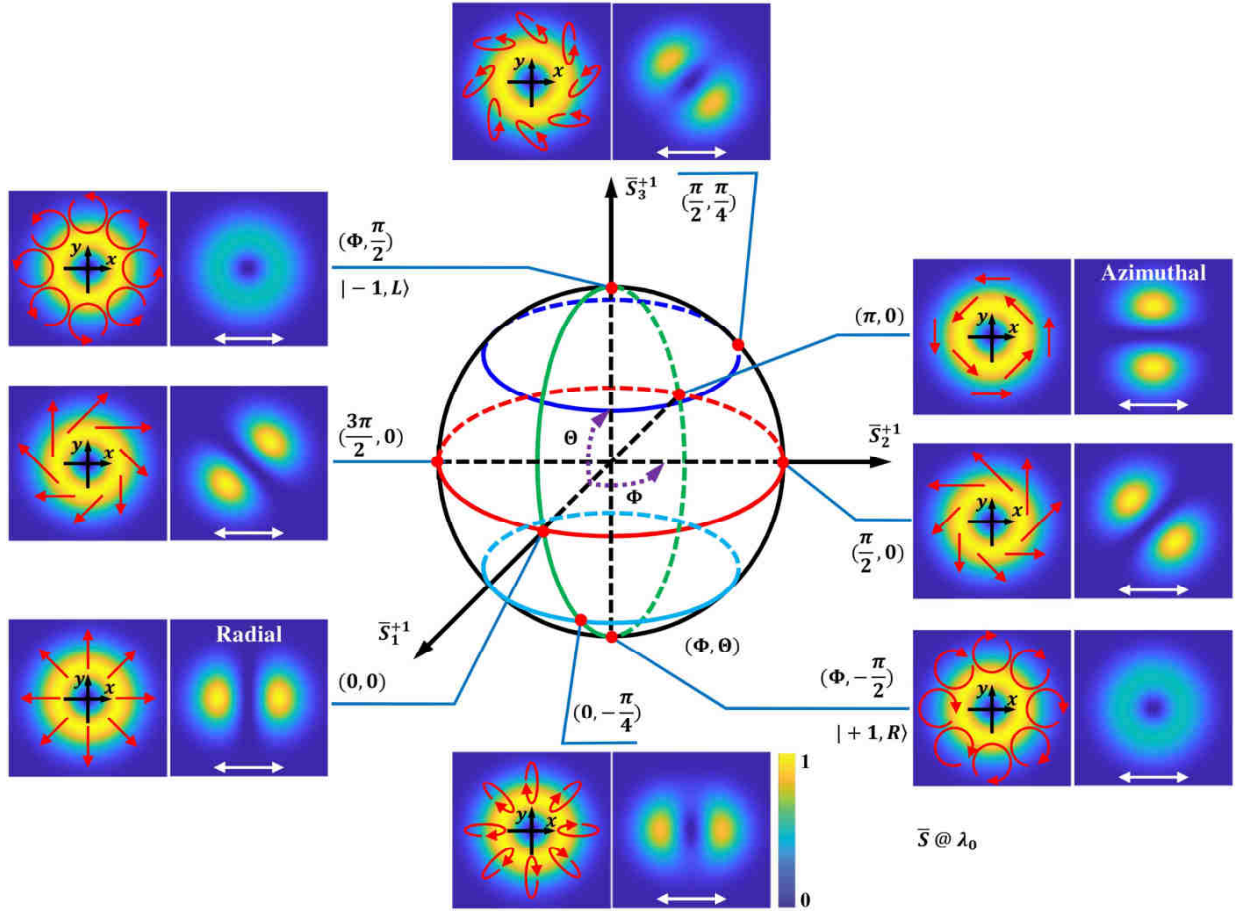

**Figure S2.** Simulations of the mean HOP sphere ($\overline{S}$) representation of vector vortex states corresponding to the center wavelength ($\lambda_0$) and five closed paths for the laser intracavity modulation on the HOP sphere. The left image of every pair of figures demonstrates the intensity pattern of the output beam cross-section, while the right image is the transmitted intensity from a linear polarizer oriented in the parallel direction, denoted by the white double-ended arrow. The red single-ended arrows show the distributions of the polarization ellipses. $\Phi$ and $\Theta$ are the polar and azimuthal angles of the spherical coordinates, respectively. $\overline{S}_i^{+1}$ ($i = 1,2,3$) denotes the three axes of the HOP sphere where +1 is the topological charge corresponding to $2q = +1$. Red circle: Path A, the equator; green circle: Path B, the circle of 0° and 180° longitude; blue circle: Path D, the circle of 45° latitude; cyan circle: Path E, the circle of −45° latitude.

**Table 2SA.** List of the theoretical (*T*) and experimental (*E*) values of the parameters and coefficients of $|-1, L\rangle$ and $|+1, R\rangle$ modulated along Path A.

| Index | $\Phi_T$ (rad.) | $\Theta_T$ (rad.) | $\alpha_T$ (rad.) | $\beta_T$ (rad.) | $\alpha_T$ (deg.) | $\beta_T$ (deg.) | Power (mW) | $c_1$ | $c_2$ |
|---|---|---|---|---|---|---|---|---|---|
| 1 | 0 | 0 | 0 | 0 | 352.0 | 355.1 | 3.493 | $\frac{1}{\sqrt{2}}$ | $\frac{1}{\sqrt{2}}$ |
| 2 | $\frac{\pi}{4}$ | 0 | $\frac{\pi}{8}$ | 0 | 14.5 | 355.1 | 3.301 | $\frac{1}{\sqrt{2}}exp(-\frac{\pi}{8}i)$ | $\frac{1}{\sqrt{2}}exp(\frac{\pi}{8}i)$ |
| 3 | $\frac{\pi}{2}$ | 0 | $\frac{\pi}{4}$ | 0 | 37.0 | 355.1 | 3.173 | $\frac{1}{\sqrt{2}}exp(-\frac{\pi}{4}i)$ | $\frac{1}{\sqrt{2}}exp(\frac{\pi}{4}i)$ |
| 4 | $\frac{3\pi}{4}$ | 0 | $\frac{3\pi}{8}$ | 0 | 59.5 | 355.1 | 3.191 | $\frac{1}{\sqrt{2}}exp(-\frac{3\pi}{8}i)$ | $\frac{1}{\sqrt{2}}exp(\frac{3\pi}{8}i)$ |
| 5 | $\pi$ | 0 | $\frac{\pi}{2}$ | 0 | 82.0 | 355.1 | 3.338 | $-\frac{1}{\sqrt{2}}i$ | $\frac{1}{\sqrt{2}}i$ |
| 6 | $\frac{5\pi}{4}$ | 0 | $\frac{5\pi}{8}$ | 0 | 104.5 | 355.1 | 3.490 | $\frac{1}{\sqrt{2}}exp(-\frac{5\pi}{8}i)$ | $\frac{1}{\sqrt{2}}exp(\frac{5\pi}{8}i)$ |
| 7 | $\frac{3\pi}{2}$ | 0 | $\frac{3\pi}{4}$ | 0 | 127.0 | 355.1 | 3.442 | $\frac{1}{\sqrt{2}}exp(-\frac{3\pi}{4}i)$ | $\frac{1}{\sqrt{2}}exp(\frac{3\pi}{4}i)$ |
| 8 | $\frac{7\pi}{4}$ | 0 | $\frac{7\pi}{8}$ | 0 | 149.5 | 355.1 | 3.347 | $\frac{1}{\sqrt{2}}exp(-\frac{7\pi}{8}i)$ | $\frac{1}{\sqrt{2}}exp(\frac{7\pi}{8}i)$ |
| 9 | $2\pi$ | 0 | $\pi$ | 0 | 172.0 | 355.1 | 3.203 | $-\frac{1}{\sqrt{2}}$ | $-\frac{1}{\sqrt{2}}$ |

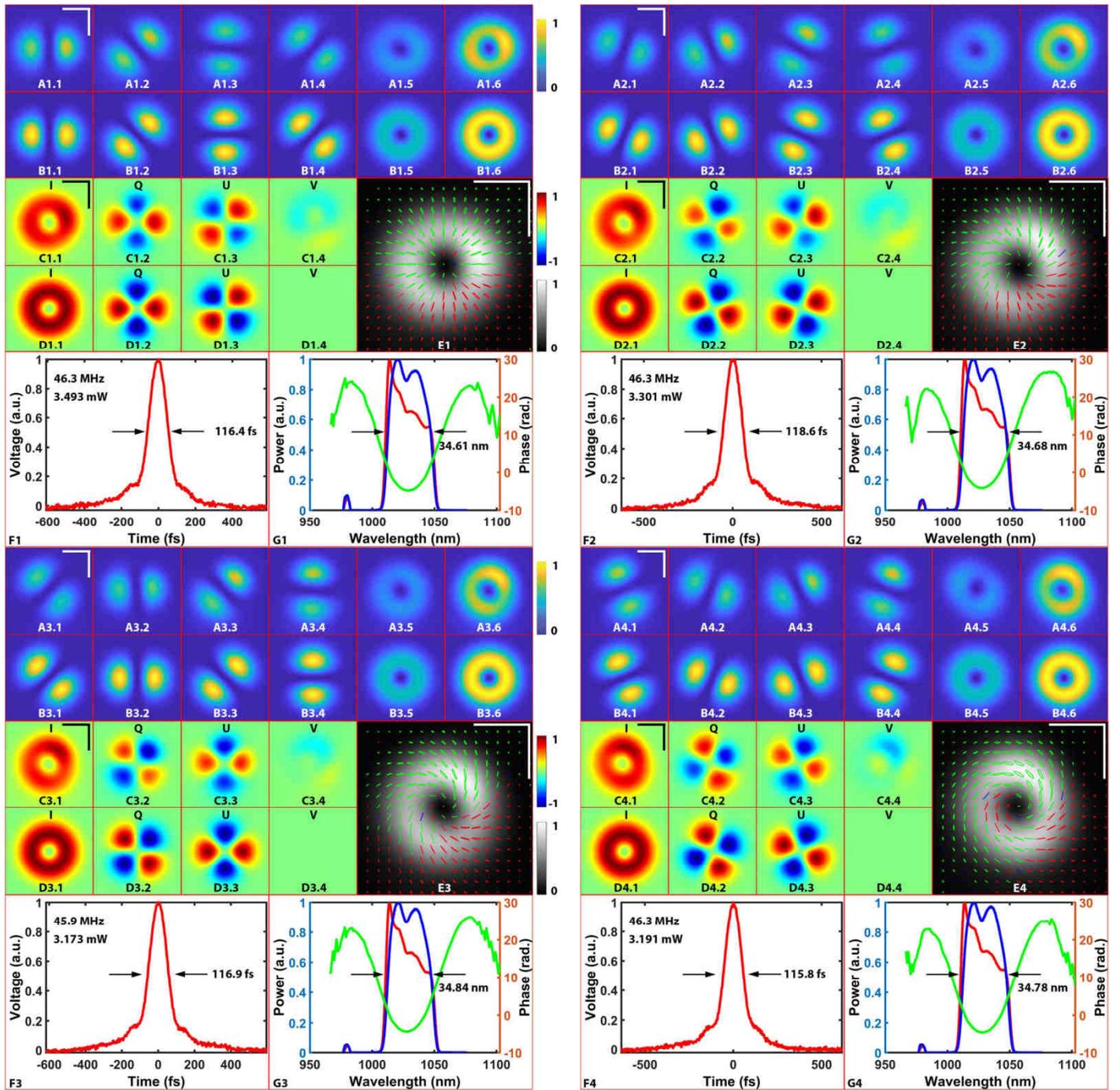

**Figure S3A1.** Conventional characterizations of the intracavity controlled OAM-carrying femtosecond laser with the states described by the (0,0), (π/4,0), (π/2,0) and (3π/4,0) on the HOP sphere $\bar{S}$. **A1-G1**: $(|-1,L\rangle + |+1,R\rangle)/\sqrt{2}$ corresponding to the radial state (0,0). **A2-G2**: $[exp\left(-\frac{\pi}{8}i\right)|-1,L\rangle + exp\left(\frac{\pi}{8}i\right)|+1,R\rangle]/\sqrt{2}$ corresponding to the state $\left(\frac{\pi}{4},0\right)$. **A3-G3**: $[exp\left(-\frac{\pi}{4}i\right)|-1,L\rangle + exp\left(\frac{\pi}{4}i\right)|+1,R\rangle]/\sqrt{2}$ corresponding to the point $\left(\frac{\pi}{2},0\right)$. **A4-G4**: $[exp\left(-\frac{3\pi}{8}i\right)|-1,L\rangle + exp\left(\frac{3\pi}{8}i\right)|+1,R\rangle]/\sqrt{2}$ corresponding to the point $\left(\frac{3\pi}{4},0\right)$. For the radial state: **A1.1**-**A1.6** are the 2D intensity measurements with different configurations. **A1.6** are the total intensity distributions; **A1.5** are the patterns after the pulses transmit through the QWP with horizontal fast axis and the polarizer with the orientation of 45°; **A1.1**-**A1.4** are the patterns when the polarizers with such orientations as: 1 → 0°, 2 → 45°, 3 → 90°, 4 → 135°; **C1.1**-**C1.4** are the Stokes parameters; **B1.1**-**B1.6**, and **D1.1**-**D1.4** are the corresponding simulations; **E1** are the polarization ellipses, where green denotes the left-handed polarization, red the right-handed polarization, and blue the linearly polarized; **F1** are the pulse duration measurements of the pulse from the inner cavity (reference beam) after the chirp compensation; **G1** is the spectrum of the pulses from the laser output coupler (blue), the reference beam (red) as well as the phase (green). The right portion of this figure shares the same setup. Scale bars represent 1 mm.

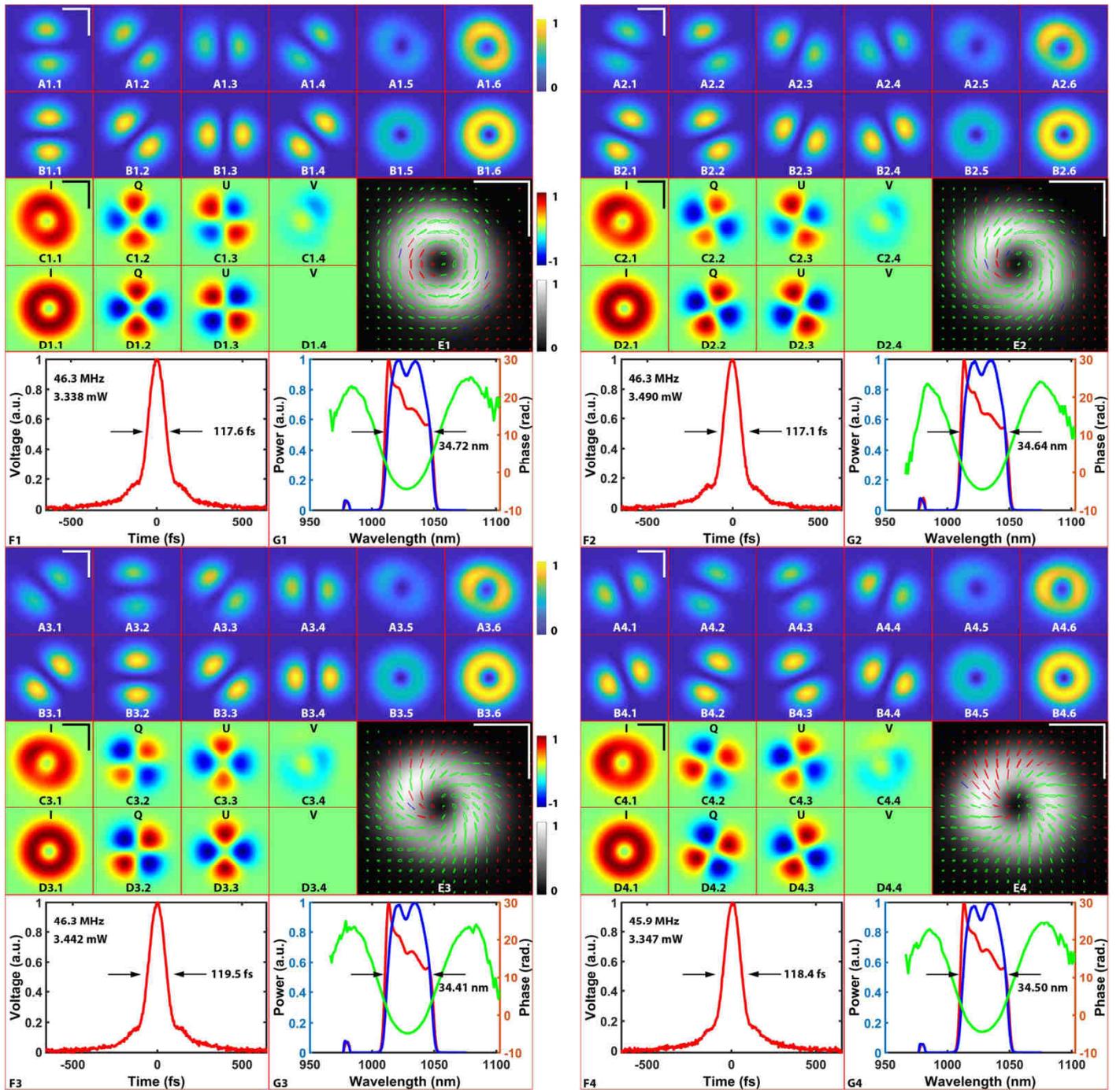

**Figure S3A2.** Conventional characterizations of the states described by the $(\pi,0)$, $(5\pi/4,0)$, $(3\pi/2,0)$, and $(7\pi/4,0)$ on $\bar{S}$. **A1-G1**: $(-|-1,L\rangle + |+1,R\rangle)i/\sqrt{2}$ corresponding to the radial state $(\pi,0)$. **A2-G2**: $[exp\left(-\frac{5\pi}{8}i\right)|-1,L\rangle + exp\left(\frac{5\pi}{8}i\right)|+1,R\rangle]/\sqrt{2}$ corresponding to the state $\left(\frac{5\pi}{4},0\right)$. **A3-G3**: $[exp\left(-\frac{3\pi}{4}i\right)|-1,L\rangle + exp\left(\frac{3\pi}{4}i\right)|+1,R\rangle]/\sqrt{2}$ corresponding to the point $\left(\frac{3\pi}{2},0\right)$. **A4-G4**: $[exp\left(-\frac{7\pi}{8}i\right)|-1,L\rangle + exp\left(\frac{7\pi}{8}i\right)|+1,R\rangle]/\sqrt{2}$ corresponding to the point $\left(\frac{7\pi}{4},0\right)$.

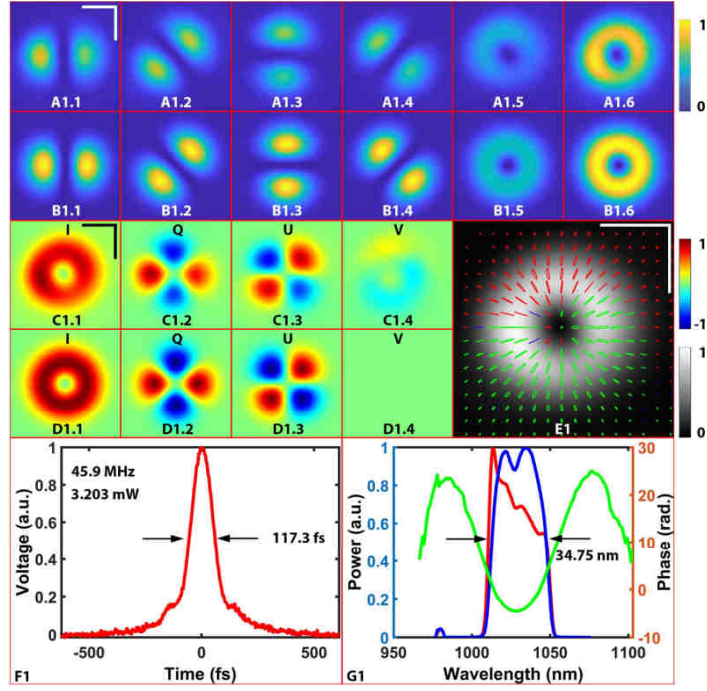

**Figure S3A3.** Conventional characterization of the states described by the $(2\pi, 0)$ on $\overline{S}$. **A1-G1**: $(|-1, L\rangle + |+1, R\rangle)/\sqrt{2}$ corresponding to the radial state $(2\pi, 0)$.

### I.3.B.  Path B, the Circle of 0° and 180° Longitude

This section presents the detailed characterization for the states modulated along the Path B. Table S2B shows the $T$ and $E$ values of $\Phi$, $\Theta$, $\alpha$, $\beta$, the output power, $c_1$ and $c_2$ for two eigenstates, $|-1, L\rangle$ and $|+1, R\rangle$. Figures S3B.1-3 are the conventional characterizations.

**Table S2B.** List of the $T$ and $E$ values of the parameters and coefficients of $|-1, L\rangle$ and $|+1, R\rangle$ modulated along Path B.

| Index | $\Phi_T$ (rad.) | $\Theta_T$ (rad.) | $\alpha_T$ (rad.) | $\beta_T$ (rad.) | $\alpha_T$ (deg.) | $\beta_T$ (deg.) | Power (mW) | $c_1$ | $c_2$ |
|---|---|---|---|---|---|---|---|---|---|
| 1 | 0 | 0 | 0 | 0 | 352 | 355.1 | 3.656 | $\frac{1}{\sqrt{2}}$ | $\frac{1}{\sqrt{2}}$ |
| 2 | 0 | $\frac{\pi}{4}$ | $\frac{\pi}{8}$ | $\frac{\pi}{8}$ | 14.5 | 332.6 | 3.348 | $\sin(\frac{3\pi}{8})$ | $\cos(\frac{3\pi}{8})$ |
| 3 | 0 | $\frac{\pi}{2}$ | $\frac{\pi}{4}$ | $\frac{\pi}{4}$ | 37 | 310.1 | 3.444 | 1 | 0 |
| 4 | $\pi$ | $\frac{\pi}{4}$ | $\frac{5\pi}{8}$ | $\frac{\pi}{8}$ | 104.5 | 332.6 | 3.474 | $-\sin(\frac{3\pi}{8})i$ | $\cos(\frac{3\pi}{8})i$ |
| 5 | $\pi$ | 0 | $\frac{\pi}{2}$ | 0 | 82 | 355.1 | 3.225 | $-\frac{1}{\sqrt{2}}i$ | $\frac{1}{\sqrt{2}}i$ |
| 6 | $\pi$ | $-\frac{\pi}{4}$ | $\frac{3\pi}{8}$ | $-\frac{\pi}{8}$ | 59.5 | 17.6 | 3.182 | $-\sin(\frac{\pi}{8})i$ | $\cos(\frac{\pi}{8})i$ |
| 7 | $\pi$ | $-\frac{\pi}{2}$ | $\frac{\pi}{4}$ | $-\frac{\pi}{4}$ | 37 | 40.1 | 3.381 | 0 | $i$ |
| 8 | 0 | $-\frac{\pi}{4}$ | $-\frac{\pi}{8}$ | $-\frac{\pi}{8}$ | 329.5 | 17.6 | 3.661 | $\sin(\frac{\pi}{8})$ | $\cos(\frac{\pi}{8})$ |
| 9 | 0 | 0 | 0 | 0 | 352 | 355.1 | 3.477 | $\frac{1}{\sqrt{2}}$ | $\frac{1}{\sqrt{2}}$ |

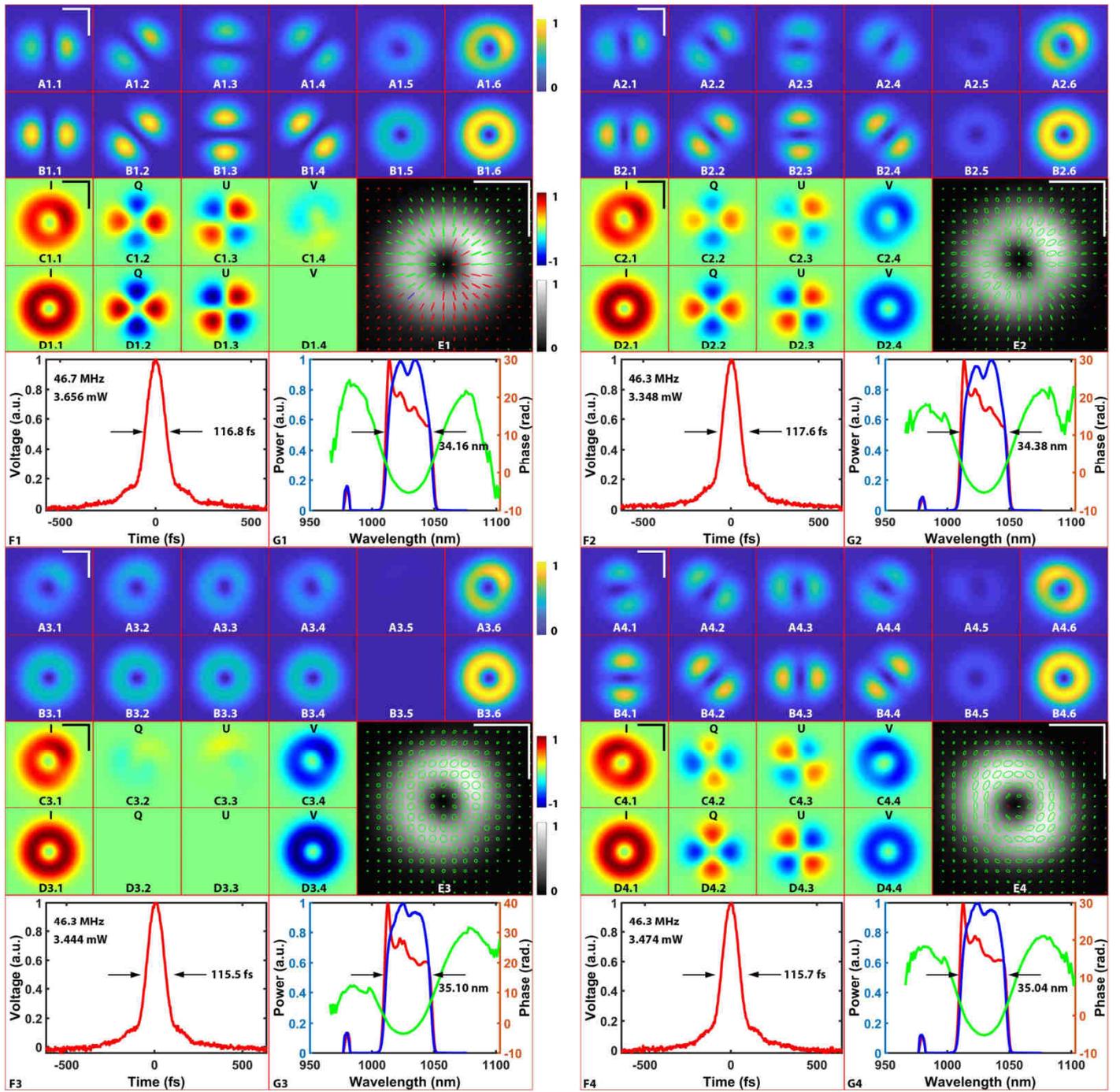

**Figure S3B1.** Conventional characterizations of the states described by $(0,0)$, $(0,\pi/4)$, $(0,\pi/2)$, and $(\pi,\pi/4)$ on $\overline{S}$. **A1-G1**: $(|-1,L\rangle + |+1,R\rangle)/\sqrt{2}$ corresponding to the radial state $(0,0)$. **A2-G2**: $[sin(\frac{3\pi}{8})|-1,L\rangle + cos(\frac{3\pi}{8})|+1,R\rangle]/\sqrt{2}$ corresponding to the state $\left(0,\frac{\pi}{4}\right)$. **A3-G3**: $[|-1,L\rangle + 0|+1,R\rangle]/\sqrt{2}$ corresponding to the point $\left(0,\frac{\pi}{2}\right)$. **A4-G4**: $[-sin(\frac{3\pi}{8})|-1,L\rangle + cos(\frac{3\pi}{8})|+1,R\rangle]i/\sqrt{2}$ corresponding to the point $(\pi,\pi/4)$.

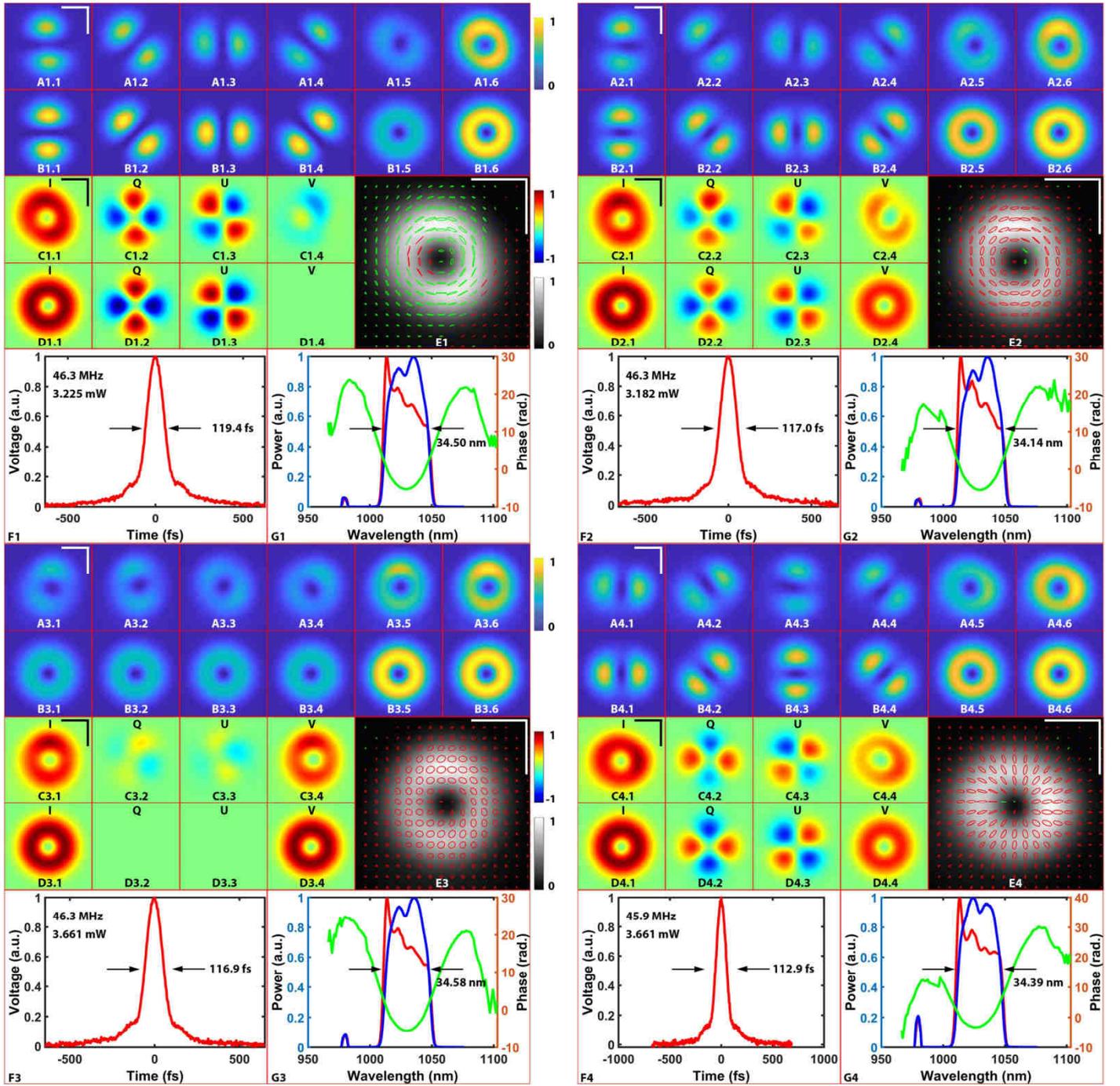

**Figure S3B2.** Conventional characterizations of the states described by $(\pi, 0)$, $(\pi, -\pi/4)$, $(\pi, -\pi/2)$, and $(0, -\pi/4)$ on $\overline{S}$. **A1-G1**: $(-|-1, L\rangle + |+1, R\rangle)i/\sqrt{2}$ corresponding to the radial state $(\pi, 0)$. **A2-G2**: $[-sin(\frac{\pi}{8})|-1, L\rangle + cos(\frac{\pi}{8})|+1, R\rangle]i/\sqrt{2}$ corresponding to the state $(\pi, -\pi/4)$. **A3-G3**: $[0|-1, L\rangle + i|+1, R\rangle]/\sqrt{2}$ corresponding to the point $(\pi, -\pi/2)$. **A4-G4**: $[sin(\frac{\pi}{8})|-1, L\rangle + cos(\frac{\pi}{8})|+1, R\rangle]/\sqrt{2}$ corresponding to the point $(0, -\pi/4)$.

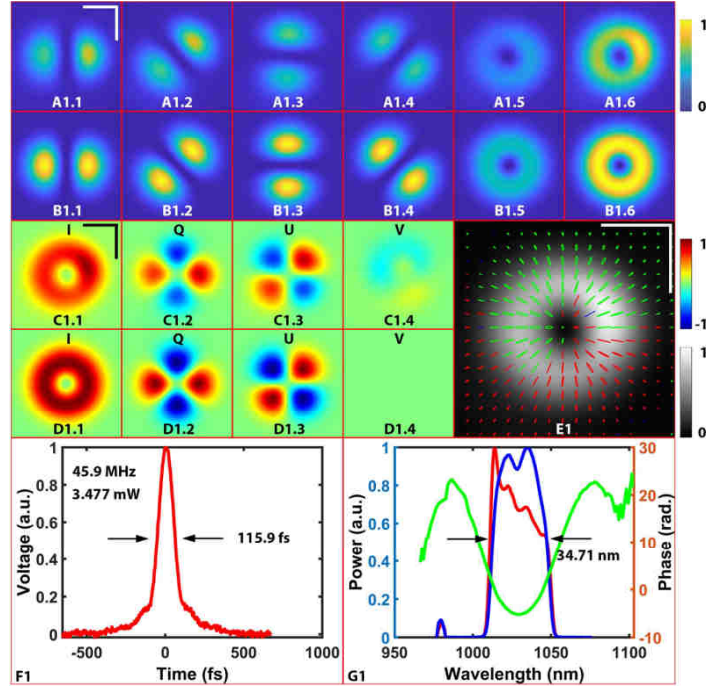

**Figure S3B3.** Conventional characterization of the state (0,0) on $\bar{S}$. **A1-G1**: $(|-1,L\rangle + |+1,R\rangle)/\sqrt{2}$ corresponding to the radial state (0,0).

### I.3.C. Path C, the Polar Triangle

This part shows the characterizations of the states modulated along Path C. Table S2C shows the $T$ and $E$ values. Figures 3C.1-3 are the conventional characterizations.

**Table S2C.** List of $T$ and $E$ values for $|-1,L\rangle$ and $|+1,R\rangle$ modulated along Path C.

| Index | $\Phi_T$ (rad.) | $\Theta_T$ (rad.) | $\alpha_T$ (rad.) | $\beta_T$ (rad.) | $\alpha_T$ (deg.) | $\beta_T$ (deg.) | Power (mW) | $c_1$ | $c_2$ |
|---|---|---|---|---|---|---|---|---|---|
| 1 | 0 | 0 | 0 | 0 | 352 | 355.1 | 3.413 | $\frac{1}{\sqrt{2}}$ | $\frac{1}{\sqrt{2}}$ |
| 2 | $\frac{\pi}{4}$ | 0 | $\frac{\pi}{8}$ | 0 | 14.5 | 355.1 | 3.194 | $\frac{1}{\sqrt{2}}exp(-\frac{\pi}{8}i)$ | $\frac{1}{\sqrt{2}}exp(\frac{\pi}{8}i)$ |
| 3 | $\frac{\pi}{2}$ | 0 | $\frac{\pi}{4}$ | 0 | 37 | 355.1 | 3.061 | $\frac{1}{\sqrt{2}}exp(-\frac{\pi}{4}i)$ | $\frac{1}{\sqrt{2}}exp(\frac{\pi}{4}i)$ |
| 4 | $\frac{\pi}{2}$ | $\frac{\pi}{4}$ | $\frac{3\pi}{8}$ | $\frac{\pi}{8}$ | 59.5 | 332.6 | 3.294 | $sin(\frac{3\pi}{8})exp(-\frac{\pi}{4}i)$ | $cos(\frac{3\pi}{8})exp(\frac{\pi}{4}i)$ |
| 5 | $\frac{\pi}{2}$ | $\frac{\pi}{2}$ | $\frac{\pi}{2}$ | $\frac{\pi}{4}$ | 82 | 310.1 | 3.675 | $exp(-\frac{\pi}{4}i)$ | 0 |
| 6 | 0 | $\frac{\pi}{2}$ | $\frac{\pi}{4}$ | $\frac{\pi}{4}$ | 37 | 310.1 | 3.257 | 1 | 0 |
| 7 | 0 | $\frac{\pi}{4}$ | $\frac{\pi}{8}$ | $\frac{\pi}{8}$ | 14.5 | 332.6 | 3.182 | $sin(\frac{3\pi}{8})$ | $cos(\frac{3\pi}{8})$ |
| 8 | 0 | 0 | 0 | 0 | 352 | 355.1 | 3.413 | $\frac{1}{\sqrt{2}}$ | $\frac{1}{\sqrt{2}}$ |

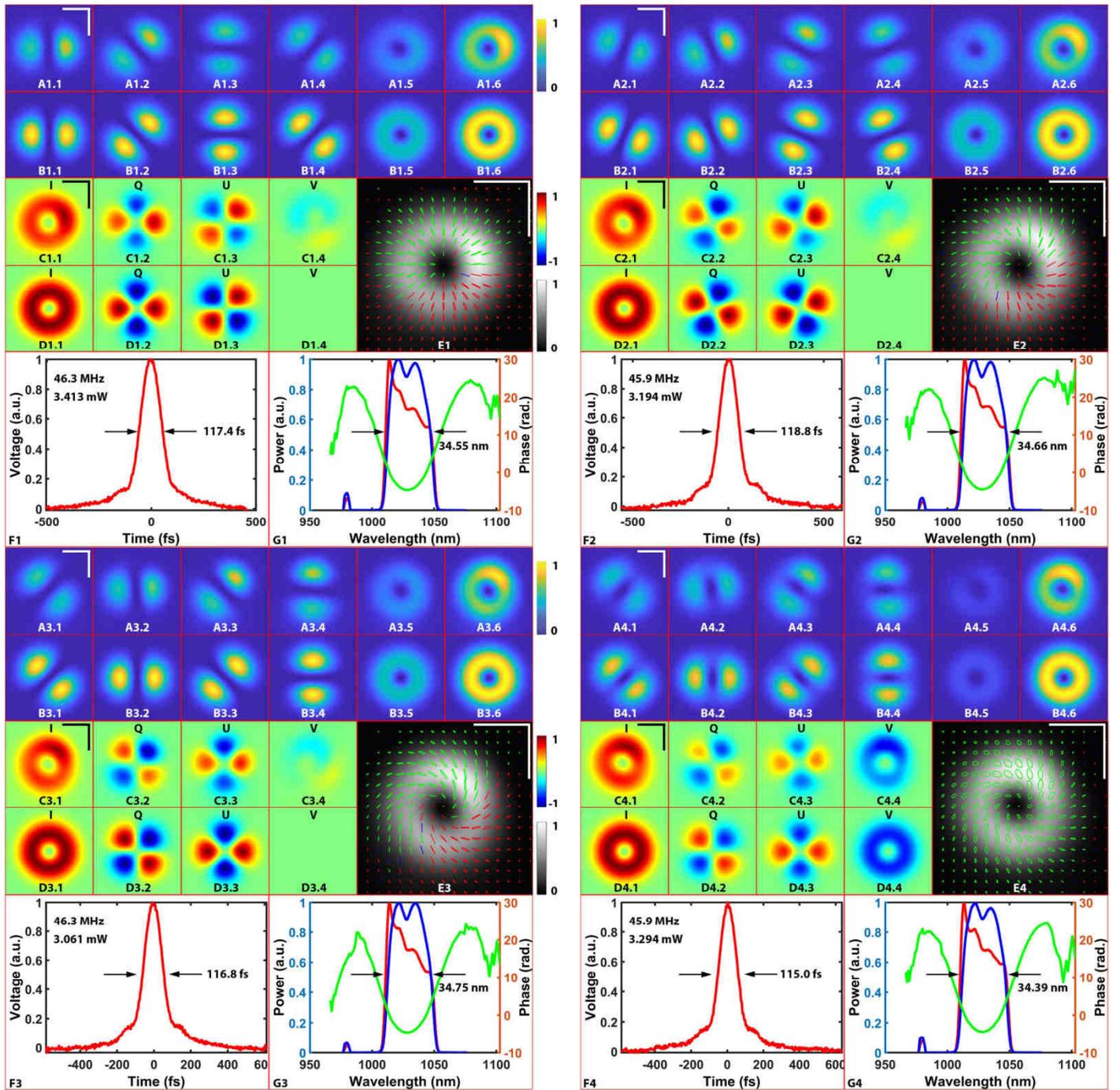

**Figure S3C1.** Conventional characterizations of the states $(0,0)$, $(\pi/4,0)$, $(\pi/2,0)$, and $(\pi/2,\pi/4)$ on $\overline{S}$. **A1-G1**: $(|-1,L\rangle+|+1,R\rangle)/\sqrt{2}$ corresponding to the radial state $(0,0)$. **A2-G2**: $[exp\left(-\frac{\pi}{8}i\right)|-1,L\rangle + exp\left(\frac{\pi}{8}i\right)|+1,R\rangle]/\sqrt{2}$ corresponding to the state $\left(\frac{\pi}{4},0\right)$. **A3-G3**: $[exp\left(-\frac{\pi}{4}i\right)|-1,L\rangle + exp\left(\frac{\pi}{4}i\right)|+1,R\rangle]/\sqrt{2}$ corresponding to the point $\left(\frac{\pi}{2},0\right)$. **A4-G4**: $[sin(\frac{3\pi}{8})\exp(-\frac{\pi}{4}i)|-1,L\rangle + cos(\frac{3\pi}{8})\exp(\frac{\pi}{4}i)|+1,R\rangle]/\sqrt{2}$ corresponding to the point $\left(\frac{\pi}{2},\frac{\pi}{4}\right)$.

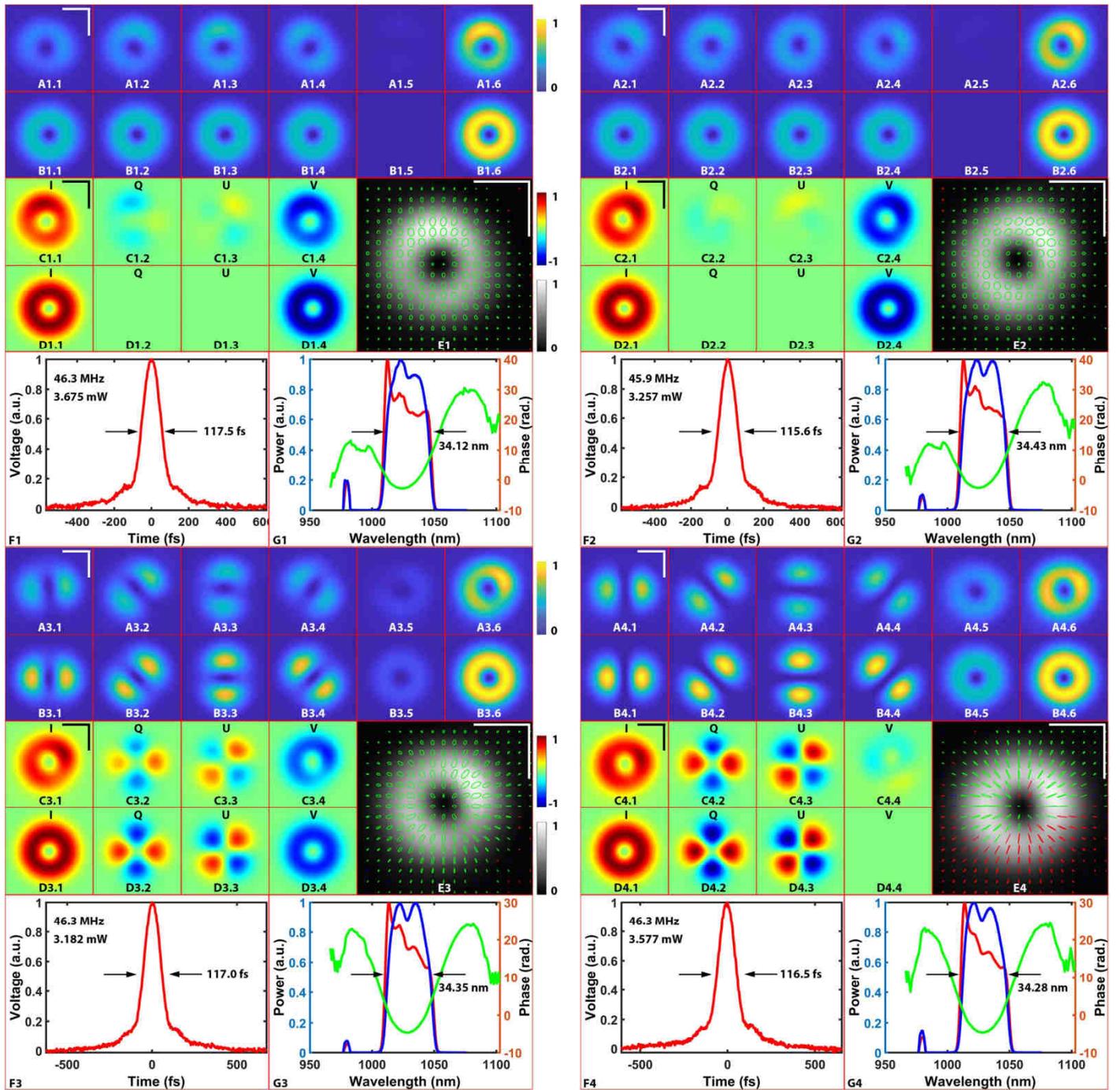

**Figure S3C2.** Conventional characterizations of the states described by the $(\pi/2, \pi/2)$, $(0, \pi/2)$, $(0, \pi/4)$, and $(0,0)$ on $\overline{S}$. **A1-G1**: $[\exp(-\frac{\pi}{4}i)|-1, L\rangle + 0|+1, R\rangle]/\sqrt{2}$ corresponding to the radial state $(\pi/2, \pi/2)$. **A2-G2**: $(1|-1, L\rangle + 0|+1, R\rangle)/\sqrt{2}$ corresponding to the state $\left(0, \frac{\pi}{2}\right)$. **A3-G3**: $[sin(\frac{3\pi}{8})|-1, L\rangle + cos(\frac{3\pi}{8})|+1, R\rangle]/\sqrt{2}$ corresponding to the point $\left(0, \frac{\pi}{4}\right)$. **A4-G4**: $(|-1, L\rangle + |+1, R\rangle)/\sqrt{2}$ corresponding to the point $(0,0)$.

## I.3.D. Path D, the Circle of 45° Latitude

This part shows the characterizations of the states modulated along the path D. Table S2D shows the $T$ and $E$ values. Figures 3D.1-3 are the conventional characterizations.

**Table S2D.** List of $T$ and $E$ values for $|-1, L\rangle$ and $|+1, R\rangle$ modulated along Path D.

| Index | $\Phi_T$ (rad.) | $\Theta_T$ (rad.) | $\alpha_T$ (rad.) | $\beta_T$ (rad.) | $\alpha_T$ (deg.) | $\beta_T$ (deg.) | Power (mW) | $c_1$ | $c_2$ |
|---|---|---|---|---|---|---|---|---|---|
| 1 | 0 | $\frac{\pi}{4}$ | $\frac{\pi}{8}$ | $\frac{\pi}{8}$ | 14.5 | 332.6 | 3.319 | $sin(\frac{3\pi}{8})$ | $cos(\frac{3\pi}{8})$ |
| 2 | $\frac{\pi}{4}$ | $\frac{\pi}{4}$ | $\frac{\pi}{4}$ | $\frac{\pi}{8}$ | 37 | 332.6 | 3.194 | $sin(\frac{3\pi}{8})exp(-\frac{\pi}{8}i)$ | $cos(\frac{3\pi}{8})exp(\frac{\pi}{8}i)$ |
| 3 | $\frac{\pi}{2}$ | $\frac{\pi}{4}$ | $\frac{3\pi}{8}$ | $\frac{\pi}{8}$ | 59.5 | 332.6 | 3.22 | $sin(\frac{3\pi}{8})exp(-\frac{\pi}{4}i)$ | $cos(\frac{3\pi}{8})exp(\frac{\pi}{4}i)$ |
| 4 | $\frac{3\pi}{4}$ | $\frac{\pi}{4}$ | $\frac{\pi}{2}$ | $\frac{\pi}{8}$ | 82 | 332.6 | 3.338 | $sin(\frac{3\pi}{8})exp(-\frac{3\pi}{8}i)$ | $cos(\frac{3\pi}{8})exp(\frac{3\pi}{8}i)$ |
| 5 | $\pi$ | $\frac{\pi}{4}$ | $\frac{5\pi}{8}$ | $\frac{\pi}{8}$ | 104.5 | 332.6 | 3.492 | $-sin(\frac{3\pi}{8})i$ | $cos(\frac{3\pi}{8})i$ |
| 6 | $\frac{5\pi}{4}$ | $\frac{\pi}{4}$ | $\frac{3\pi}{4}$ | $\frac{\pi}{8}$ | 127 | 332.6 | 3.592 | $sin(\frac{3\pi}{8})exp(-\frac{5\pi}{8}i)$ | $cos(\frac{3\pi}{8})exp(\frac{5\pi}{8}i)$ |
| 7 | $\frac{3\pi}{2}$ | $\frac{\pi}{4}$ | $\frac{7\pi}{8}$ | $\frac{\pi}{8}$ | 149.5 | 332.6 | 3.542 | $sin(\frac{3\pi}{8})exp(-\frac{3\pi}{4}i)$ | $cos(\frac{3\pi}{8})exp(\frac{3\pi}{4}i)$ |
| 8 | $\frac{7\pi}{4}$ | $\frac{\pi}{4}$ | $\pi$ | $\frac{\pi}{8}$ | 172 | 332.6 | 3.446 | $sin(\frac{3\pi}{8})exp(-\frac{7\pi}{8}i)$ | $cos(\frac{3\pi}{8})exp(\frac{7\pi}{8}i)$ |
| 9 | $2\pi$ | $\frac{\pi}{4}$ | $\frac{9\pi}{8}$ | $\frac{\pi}{8}$ | 194.5 | 332.6 | 3.379 | $-sin(\frac{3\pi}{8})$ | $-cos(\frac{3\pi}{8})$ |
| 10 | $2\pi$ | 0 | 0 | 0 | 352 | 355.1 | 3.51 | $-\frac{1}{\sqrt{2}}$ | $-\frac{1}{\sqrt{2}}$ |

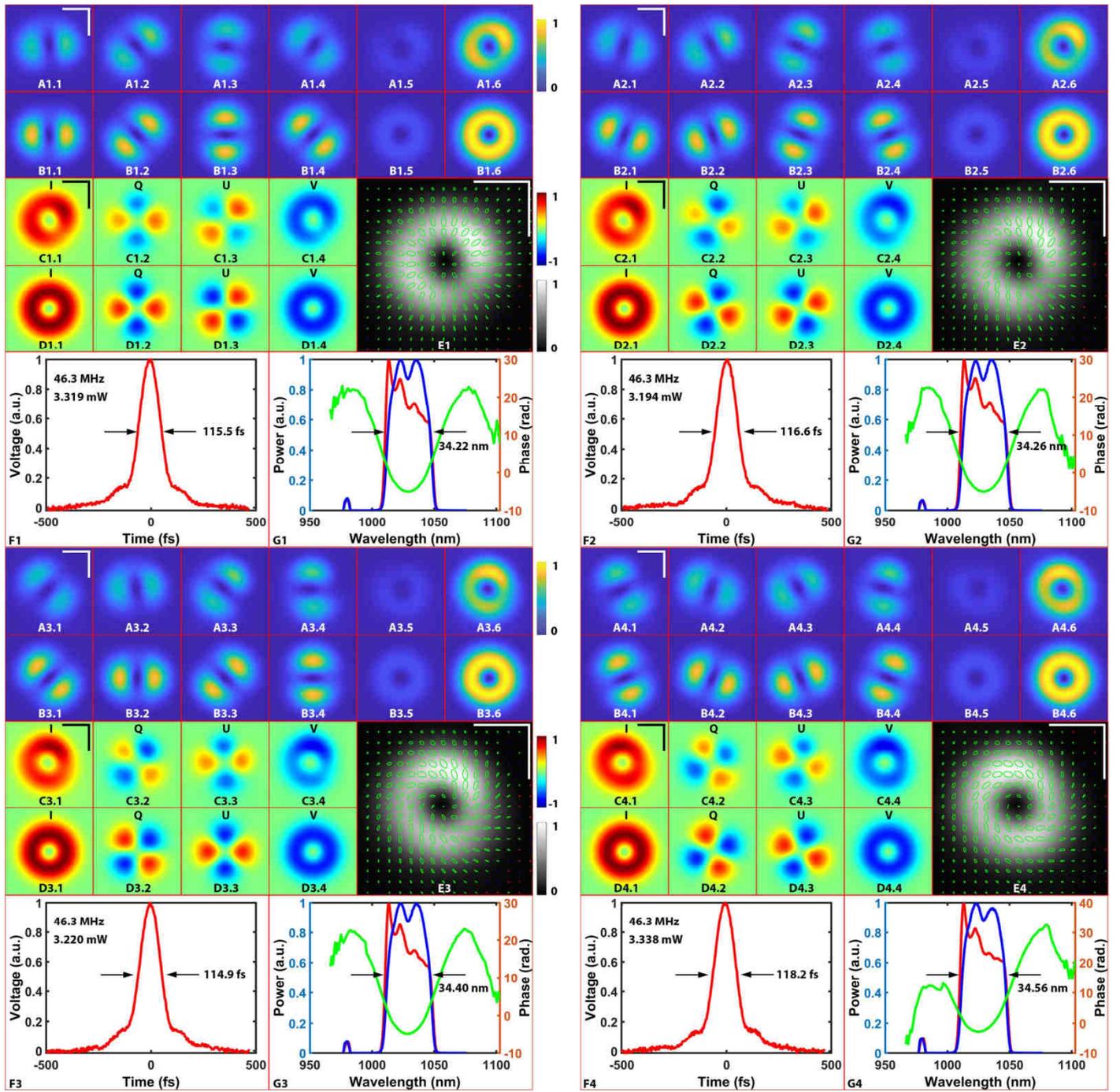

**Figure S3D1.** Conventional characterizations of the states $(0, \pi/4)$, $(\pi/4, \pi/4)$, $(\pi/2, \pi/4)$, and $(3\pi/4, \pi/4)$ on $\overline{S}$. **A1-G1**: $(sin(\frac{3\pi}{8})|-1, L\rangle + cos(\frac{3\pi}{8})|+1, R\rangle)/\sqrt{2}$ corresponding to the radial state $(0, \pi/4)$. **A2-G2**: $[sin(\frac{3\pi}{8})\exp(-\frac{\pi}{8}i)|-1, L\rangle + cos(\frac{3\pi}{8})\exp(\frac{\pi}{8}i)|+1, R\rangle]/\sqrt{2}$ corresponding to the state $(\pi/4, \pi/4)$. **A3-G3**: $[sin(\frac{3\pi}{8})\exp(-\frac{\pi}{4}i)|-1, L\rangle + cos(\frac{3\pi}{8})\exp(\frac{\pi}{4}i)|+1, R\rangle]/\sqrt{2}$ corresponding to the point $(\pi/2, \pi/4)$. **A4-G4**: $[sin(\frac{3\pi}{8})\exp(-\frac{3\pi}{8}i)|-1, L\rangle + cos(\frac{3\pi}{8})\exp(\frac{3\pi}{8}i)|+1, R\rangle]/\sqrt{2}$ corresponding to the point $(3\pi/4, \pi/4)$.

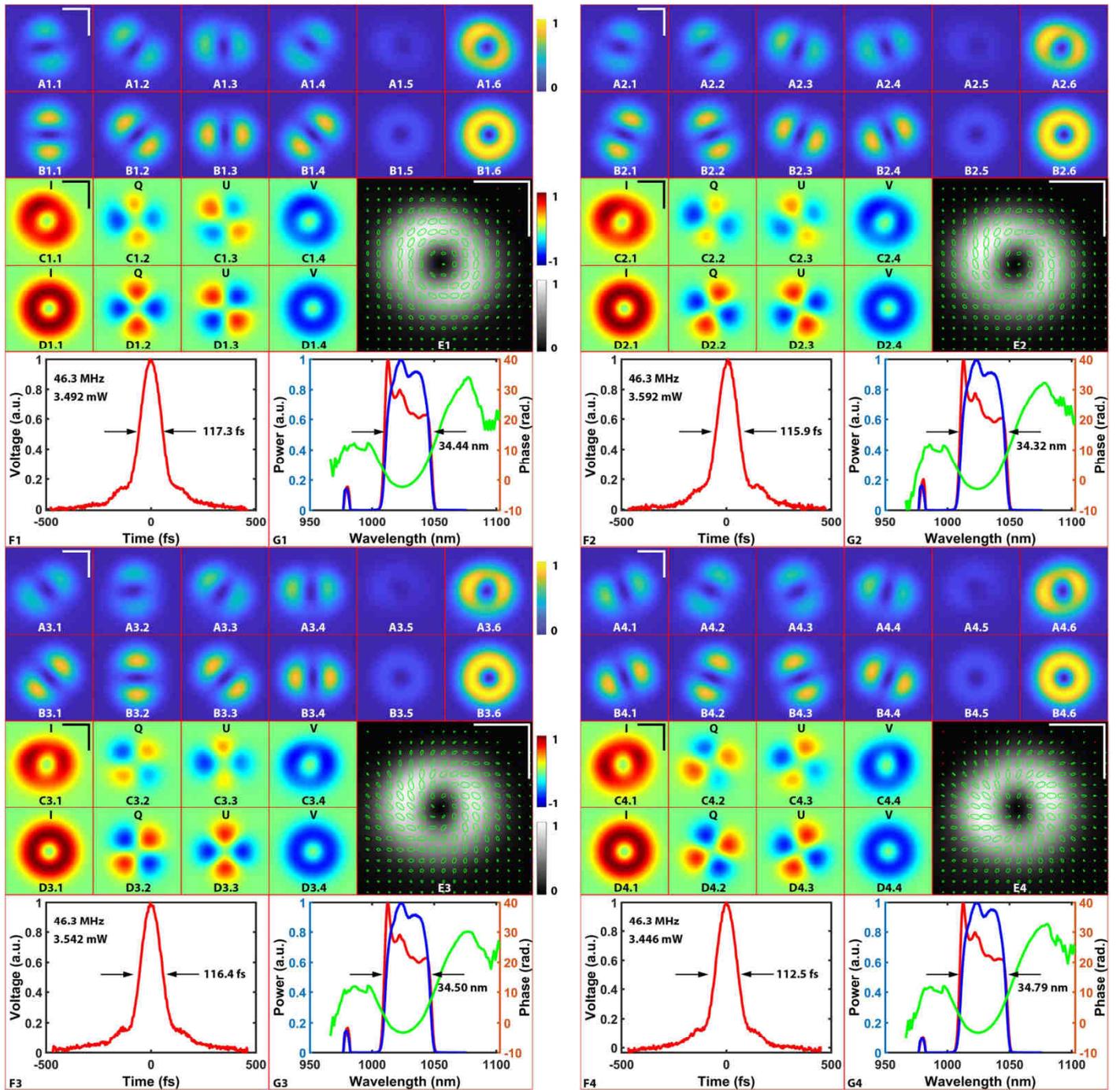

**Figure S3D2.** Conventional characterizations of the states $(\pi, \pi/4)$, $(5\pi/4, \pi/4)$, $(3\pi/2, \pi/4)$, and $(7\pi/4, \pi/4)$ on $\overline{S}$. **A1-G1**: $[-sin\left(\frac{3\pi}{8}\right)|-1,L\rangle + cos\left(\frac{3\pi}{8}\right)|+1,R\rangle]i/\sqrt{2}$ corresponding to the radial state $(\pi, \pi/4)$. **A2-G2**: $[sin(\frac{3\pi}{8})\exp(-\frac{5\pi}{8}i)|-1,L\rangle + cos(\frac{3\pi}{8})\exp(\frac{5\pi}{8}i)|+1,R\rangle]/\sqrt{2}$ corresponding to the state $(5\pi/4, \pi/4)$. **A3-G3**: $[sin(\frac{3\pi}{8})\exp(-\frac{3\pi}{4}i)|-1,L\rangle + cos(\frac{3\pi}{8})\exp(\frac{3\pi}{4}i)|+1,R\rangle]/\sqrt{2}$ corresponding to the point $(3\pi/2, \pi/4)$. **A4-G4**: $[sin(\frac{3\pi}{8})\exp(-\frac{7\pi}{8}i)|-1,L\rangle + cos(\frac{3\pi}{8})\exp(\frac{7\pi}{8}i)|+1,R\rangle]/\sqrt{2}$ corresponding to the point $(7\pi/4, \pi/4)$.

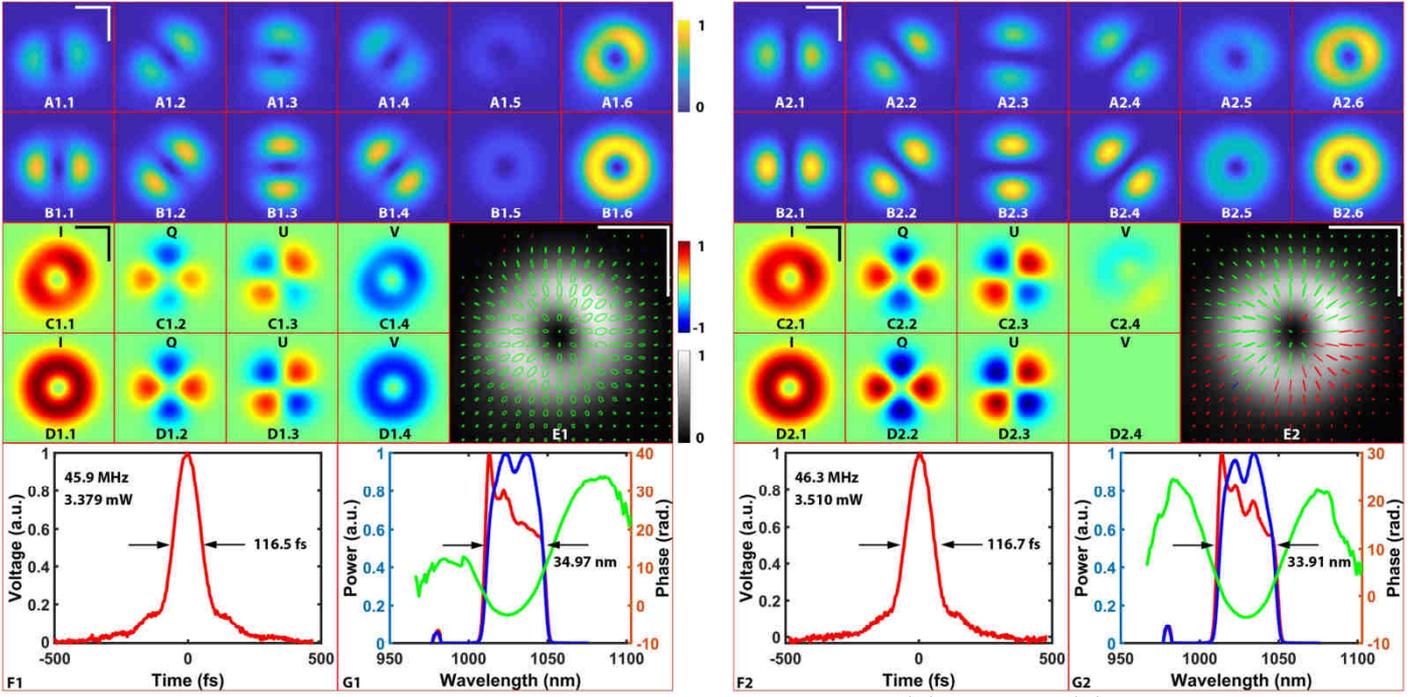

**Figure S3D3.** Conventional characterizations of the states $(2\pi, \pi/4)$ and $(2\pi, 0)$ on $\overline{S}$. **A1-G1**: $-[sin\left(\frac{3\pi}{8}\right)|-1, L\rangle + cos\left(\frac{3\pi}{8}\right)|+1, R\rangle]/\sqrt{2}$ corresponding to the radial state $(2\pi, \pi/4)$. **A2-G2**: $-(|-1, L\rangle + |+1, R\rangle)/\sqrt{2}$ corresponding to the state $(2\pi, 0)$.

## I.3.E. Path E, the Circle of −45° Latitude

This part demonstrates the characterizations of the states modulated along Path E. Table S2E shows the $T$ and $E$ values, and Figures 3E.1-3 are the conventional characterizations.

| Index | $\Phi_T$ (rad.) | $\Theta_T$ (rad.) | $\alpha_T$ (rad.) | $\beta_T$ (rad.) | $\alpha_T$ (deg.) | $\beta_T$ (deg.) | Power (mW) | $c_1$ | $c_2$ |
|---|---|---|---|---|---|---|---|---|---|
| 1 | 0 | $-\frac{\pi}{4}$ | $-\frac{\pi}{8}$ | $-\frac{\pi}{8}$ | 329.5 | 17.6 | 3.319 | $sin(\frac{\pi}{8})$ | $cos(\frac{\pi}{8})$ |
| 2 | $\frac{\pi}{4}$ | $-\frac{\pi}{4}$ | 0 | $-\frac{\pi}{8}$ | 352 | 17.6 | 3.194 | $sin(\frac{\pi}{8})\exp(-\frac{\pi}{8}i)$ | $cos(\frac{\pi}{8})\exp(\frac{\pi}{8}i)$ |
| 3 | $\frac{\pi}{2}$ | $-\frac{\pi}{4}$ | $\frac{\pi}{8}$ | $-\frac{\pi}{8}$ | 14.5 | 17.6 | 3.22 | $sin(\frac{\pi}{8})\exp(-\frac{\pi}{4}i)$ | $cos(\frac{\pi}{8})\exp(\frac{\pi}{4}i)$ |
| 4 | $\frac{3\pi}{4}$ | $-\frac{\pi}{4}$ | $\frac{\pi}{4}$ | $-\frac{\pi}{8}$ | 37 | 17.6 | 3.338 | $sin(\frac{\pi}{8})\exp(-\frac{3\pi}{8}i)$ | $cos(\frac{\pi}{8})\exp(\frac{3\pi}{8}i)$ |
| 5 | $\pi$ | $-\frac{\pi}{4}$ | $\frac{3\pi}{8}$ | $-\frac{\pi}{8}$ | 59.5 | 17.6 | 3.492 | $-sin(\frac{\pi}{8})i$ | $cos(\frac{\pi}{8})i$ |
| 6 | $\frac{5\pi}{4}$ | $-\frac{\pi}{4}$ | $\frac{\pi}{2}$ | $-\frac{\pi}{8}$ | 82 | 17.6 | 3.592 | $sin(\frac{\pi}{8})\exp(-\frac{5\pi}{8}i)$ | $cos(\frac{\pi}{8})\exp(\frac{5\pi}{8}i)$ |
| 7 | $\frac{3\pi}{2}$ | $-\frac{\pi}{4}$ | $\frac{5\pi}{8}$ | $-\frac{\pi}{8}$ | 104.5 | 17.6 | 3.542 | $sin(\frac{\pi}{8})\exp(-\frac{3\pi}{4}i)$ | $cos(\frac{\pi}{8})\exp(\frac{3\pi}{4}i)$ |
| 8 | $\frac{7\pi}{4}$ | $-\frac{\pi}{4}$ | $\frac{3\pi}{4}$ | $-\frac{\pi}{8}$ | 127 | 17.6 | 3.446 | $sin(\frac{\pi}{8})\exp(-\frac{7\pi}{8}i)$ | $cos(\frac{\pi}{8})\exp(\frac{7\pi}{8}i)$ |
| 9 | $2\pi$ | $-\frac{\pi}{4}$ | $\frac{7\pi}{8}$ | $-\frac{\pi}{8}$ | 149.5 | 17.6 | 3.379 | $-sin(\frac{\pi}{8})$ | $-cos(\frac{\pi}{8})$ |
| 10 | $2\pi$ | 0 | 0 | 0 | 352 | 355.1 | 3.51 | $-\frac{1}{\sqrt{2}}$ | $-\frac{1}{\sqrt{2}}$ |

**Table S2E.** List of $T$ and $E$ values for $|-1, L\rangle$ and $|+1, R\rangle$ modulated along Path E.

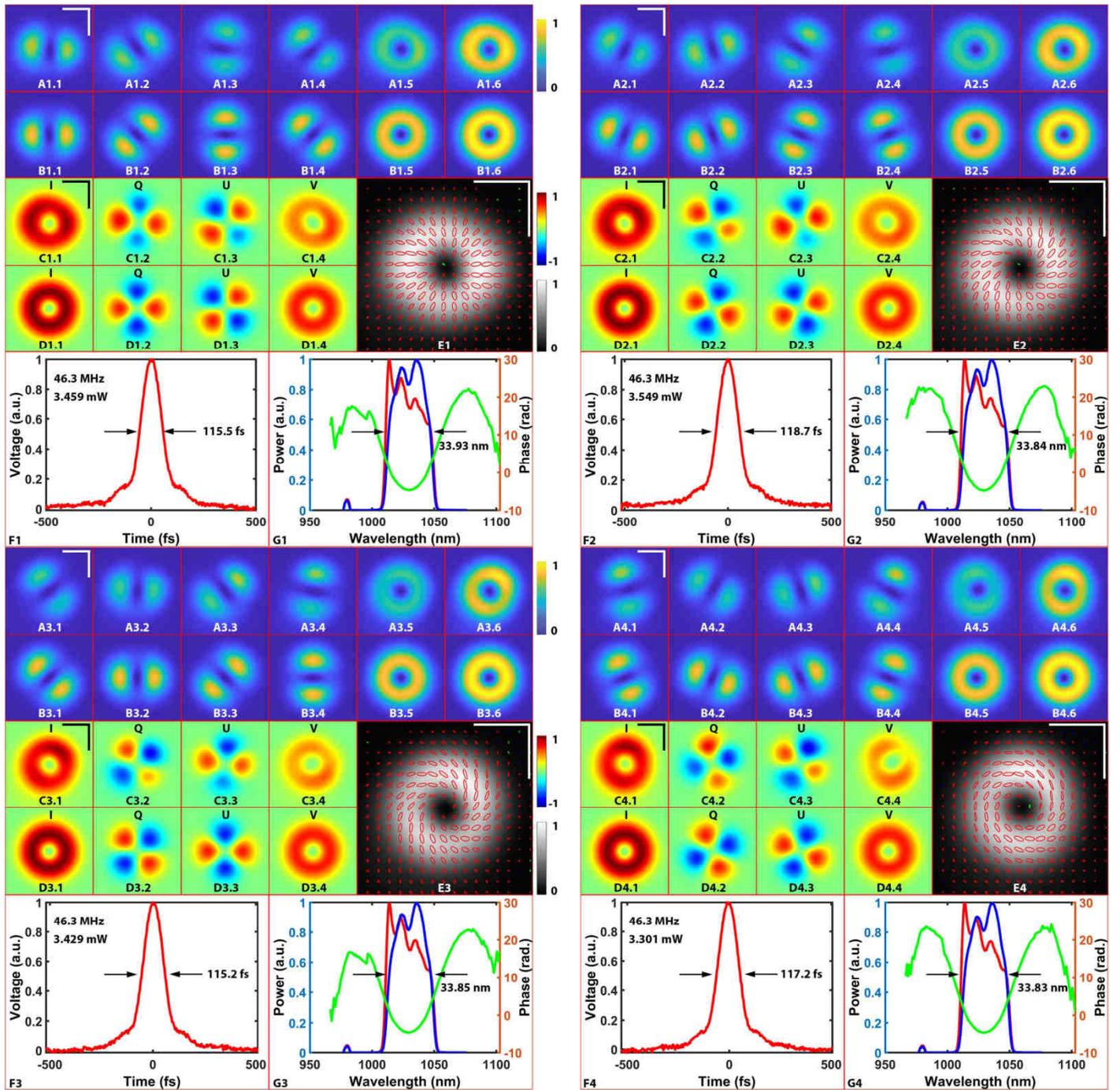

**Figure S3E1.** Conventional characterizations of the states $(0, -\pi/4)$, $(\pi/4, -\pi/4)$, $(\pi/2, -\pi/4)$, and $(3\pi/4, -\pi/4)$ on $\overline{S}$. **A1-G1**: $[sin(\frac{\pi}{8})|-1, L\rangle + cos(\frac{\pi}{8})|+1, R\rangle]/\sqrt{2}$ corresponding to the radial state $(0, -\pi/4)$. **A2-G2**: $[sin(\frac{\pi}{8})\exp(-\frac{\pi}{8}i)|-1, L\rangle + cos(\frac{\pi}{8})\exp(\frac{\pi}{8}i)|+1, R\rangle]/\sqrt{2}$ corresponding to the state $(\pi/4, -\pi/4)$. **A3-G3**: $[sin(\frac{\pi}{8})\exp(-\frac{\pi}{4}i)|-1, L\rangle + cos(\frac{\pi}{8})\exp(\frac{\pi}{4}i)|+1, R\rangle]/\sqrt{2}$ corresponding to the point $(\pi/2, -\pi/4)$. **A4-G4**: $[sin(\frac{\pi}{8})\exp(-\frac{3\pi}{8}i)|-1, L\rangle + cos(\frac{\pi}{8})\exp(\frac{3\pi}{8}i)|+1, R\rangle]/\sqrt{2}$ corresponding to the point $(3\pi/4, -\pi/4)$.

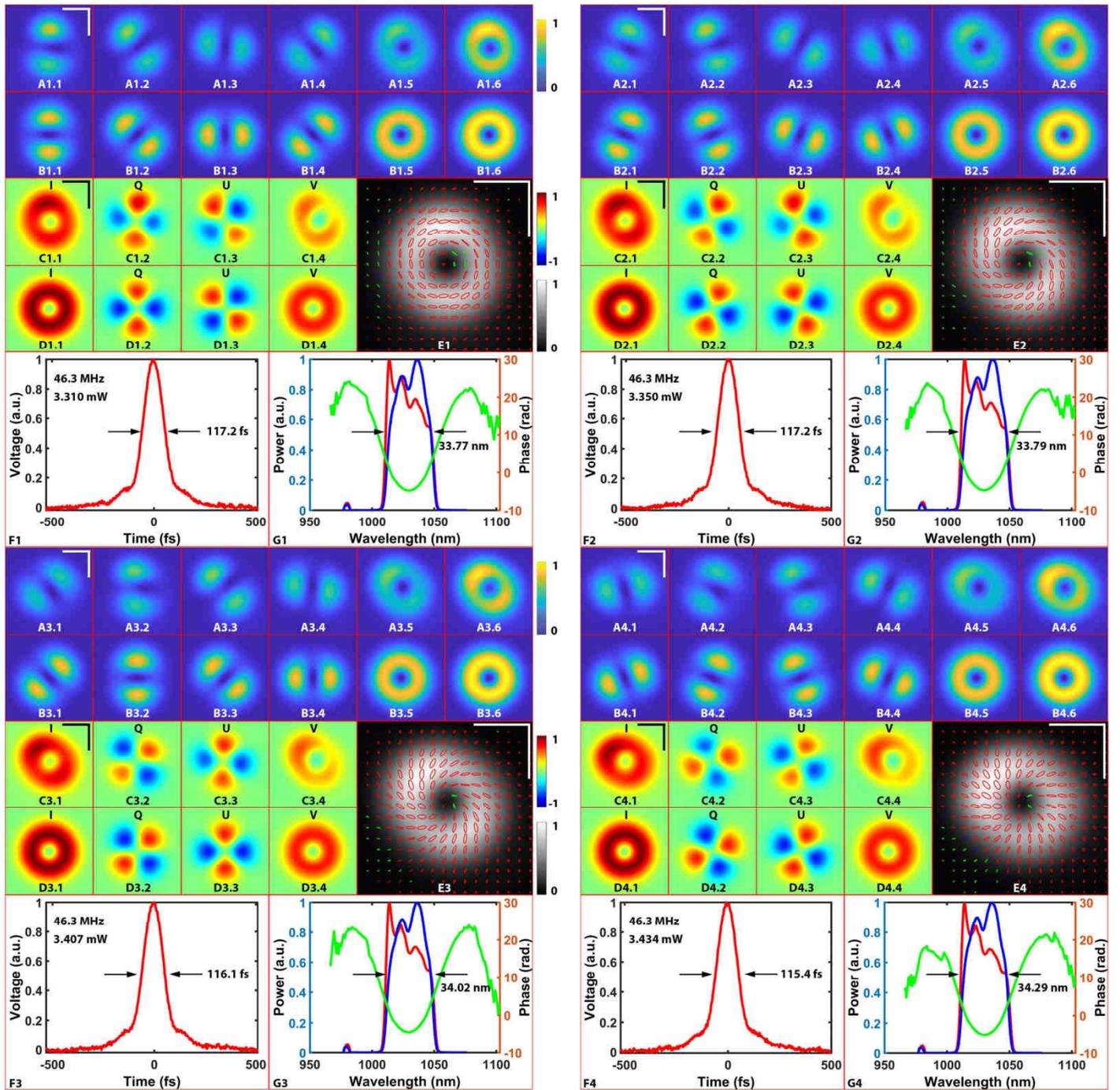

**Figure S3E2.** Conventional characterizations of the states described by $(\pi, -\pi/4)$, $\left(\frac{5\pi}{4}, -\pi/4\right)$, $\left(\frac{3\pi}{2}, -\pi/4\right)$, and $\left(\frac{7\pi}{4}, -\pi/4\right)$ on $\overline{S}$. **A1-G1**: $[-sin\left(\frac{\pi}{8}\right)|-1, L\rangle + cos\left(\frac{\pi}{8}\right)|+1, R\rangle]i/\sqrt{2}$ corresponding to the radial state $\left(\pi, -\frac{\pi}{4}\right)$. **A2-G2**: $[sin(\frac{\pi}{8})\exp(-\frac{5\pi}{8}i)|-1, L\rangle + cos(\frac{\pi}{8})\exp(\frac{5\pi}{8}i)|+1, R\rangle]/\sqrt{2}$ corresponding to the state $\left(\frac{5\pi}{4}, -\pi/4\right)$. **A3-G3**: $[sin(\frac{\pi}{8})\exp(-\frac{3\pi}{4}i)|-1, L\rangle + cos(\frac{\pi}{8})\exp(\frac{3\pi}{4}i)|+1, R\rangle]/\sqrt{2}$ corresponding to the point $\left(\frac{3\pi}{2}, -\pi/4\right)$. **A4-G4**: $[sin(\frac{\pi}{8})\exp(-\frac{7\pi}{8}i)|-1, L\rangle + cos(\frac{\pi}{8})\exp(\frac{7\pi}{8}i)|+1, R\rangle]/\sqrt{2}$ corresponding to the point $\left(\frac{7\pi}{4}, -\pi/4\right)$.

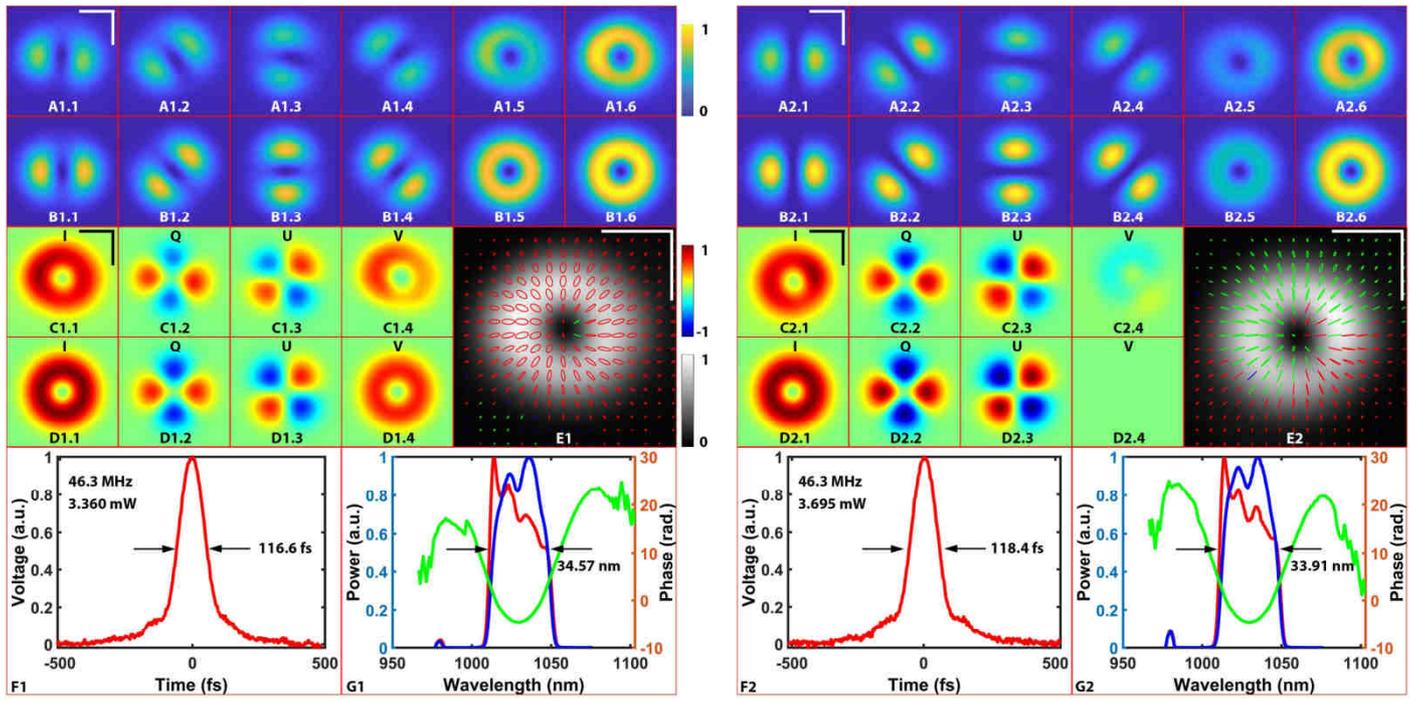

**Figure S3E3.** Conventional characterizations of the states $(2\pi, -\pi/4)$ and $(2\pi, 0)$ on $\overline{S}$. **A1-G1**: $-[sin\left(\frac{\pi}{8}\right)|-1, L\rangle + cos\left(\frac{\pi}{8}\right)|+1, R\rangle]/\sqrt{2}$ corresponding to the radial state $(2\pi, \pi/4)$. **A2-G2**: $-(|-1, L\rangle + |+1, R\rangle)/\sqrt{2}$ corresponding to the state $(2\pi, 0)$.

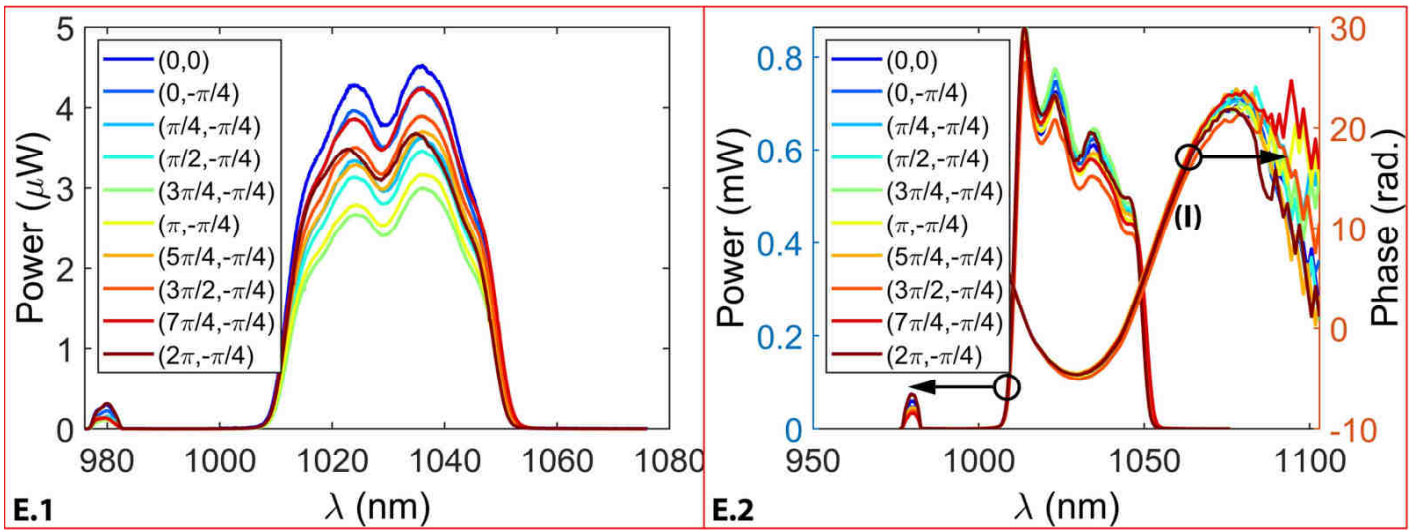

**Figure S3F.** Spectrum and phase measurements of the pulse continually modulated along Path E on the HOP sphere. Every spherical coordinate ($\Phi, \Theta$) in the legend denotes a measurement point. Roman numeral I denotes Zone I.

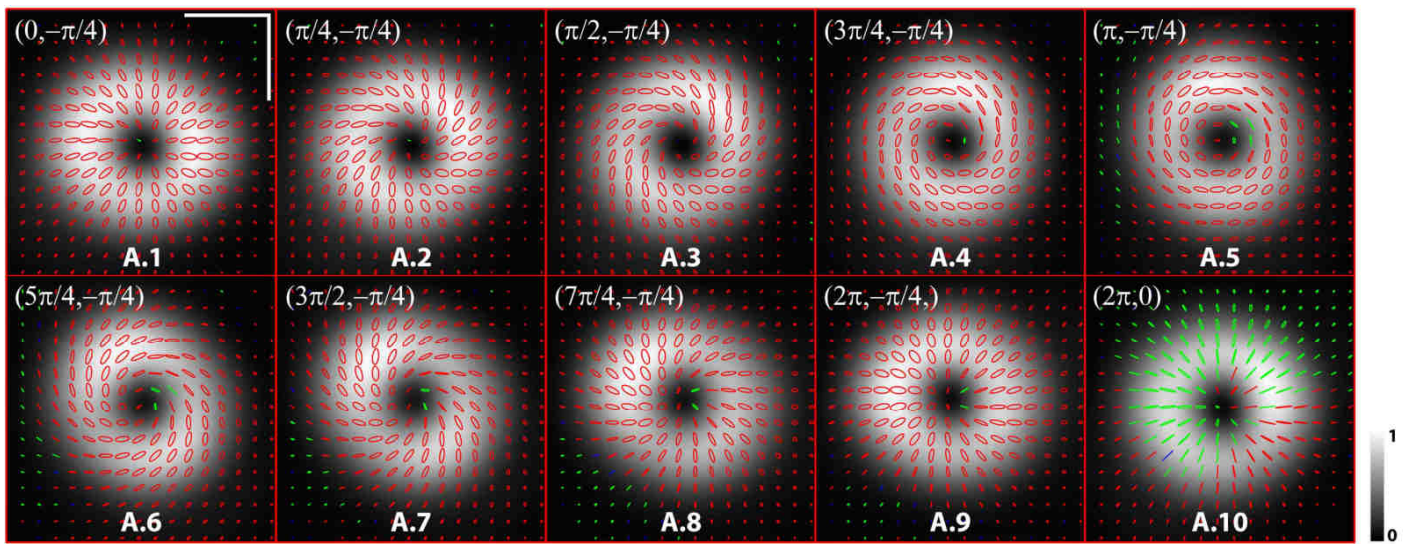

**Figure S3G.** The polarization ellipses of a pulse continually modulated along Path E on the HOP sphere. Green denotes the left-handed polarization, red the right-handed polarization, and blue the linearly polarized. The radial point state was used as the starting point, hence the 10 measurement points in Path E. Scale bar represents 1 mm.

**Section II: Detailed numerical processing for obtaining the spatiospectral and spatiotemporal properties of the femtosecond pulses**

**I.1. The principles of the spatiotemporal pulse characterization**

On plane $\Pi$, the time domain and spectral domain electric field of the reference beam can be denoted by a three components matrix:

$$\boldsymbol{E}^{ref}(\boldsymbol{r},t) = \begin{pmatrix} E_x^{ref}(\boldsymbol{r},t) \\ E_y^{ref}(\boldsymbol{r},t) \\ 0 \end{pmatrix}, \text{ and } \tilde{\boldsymbol{E}}^{ref}(\boldsymbol{r},\omega) = \begin{pmatrix} \tilde{E}_x^{ref}(\boldsymbol{r},\omega) \\ \tilde{E}_y^{ref}(\boldsymbol{r},\omega) \\ 0 \end{pmatrix} \quad \text{(S7)}$$

and for the sample beam:

$$\boldsymbol{E}^{sam}(\boldsymbol{r},t) = \begin{pmatrix} E_x^{sam}(\boldsymbol{r},t) \\ E_y^{sam}(\boldsymbol{r},t) \\ E_z^{sam}(\boldsymbol{r},t) \end{pmatrix}, \text{ and } \tilde{\boldsymbol{E}}^{sam}(\boldsymbol{r},\omega) = \begin{pmatrix} \tilde{E}_x^{sam}(\boldsymbol{r},\omega) \\ \tilde{E}_y^{sam}(\boldsymbol{r},\omega) \\ \tilde{E}_z^{sam}(\boldsymbol{r},\omega) \end{pmatrix} \quad \text{(S8)}$$

Based on the semiclassical theory of photoelectric detection of light, the averaged intensity at point $\boldsymbol{r}$ on $\Pi$ becomes, under the assumption of stationarity and ergodicity:

$$I(\boldsymbol{r},\tau) = \langle (\boldsymbol{E}^{sam}(\boldsymbol{r},t) + \boldsymbol{E}^{ref}(\boldsymbol{r},t+\tau))^* \cdot (\boldsymbol{E}^{sam}(\boldsymbol{r},t) + \boldsymbol{E}^{ref}(\boldsymbol{r},t+\tau)) \rangle$$

$$= \lim_{T \to \infty} \frac{1}{2T} \int_{-T}^{T} \left(\boldsymbol{E}^{sam}(\boldsymbol{r},t) + \boldsymbol{E}^{ref}(\boldsymbol{r},t+\tau)\right)^* \cdot \left(\boldsymbol{E}^{sam}(\boldsymbol{r},t) + \boldsymbol{E}^{ref}(\boldsymbol{r},t+\tau)\right) dt$$

$$= \langle |\boldsymbol{E}^{sam}(\boldsymbol{r},t)|^2 \rangle + \langle |\boldsymbol{E}^{ref}(\boldsymbol{r},t)|^2 \rangle + 2Re(\langle (\boldsymbol{E}^{sam}(\boldsymbol{r},t))^* \cdot (\boldsymbol{E}^{ref}(\boldsymbol{r},t+\tau)) \rangle) \quad \text{(S9)}$$

By applying the Fourier transformation ($\mathcal{F}$), Eq. S6 can be separated into 3 terms:

$$\tilde{I}(\boldsymbol{r},\omega) = \mathcal{F}\left[\langle |\boldsymbol{E}^{sam}(\boldsymbol{r},t)|^2 \rangle + \langle |\boldsymbol{E}^{ref}(\boldsymbol{r},t)|^2 \rangle\right] + \left(\tilde{\boldsymbol{E}}^{sam}(\boldsymbol{r},\omega)\right)^* \cdot \left(\tilde{\boldsymbol{E}}^{ref}(\boldsymbol{r},\omega)\right) + (\tilde{\boldsymbol{E}}^{ref}(\boldsymbol{r},-\omega))^* \cdot (\tilde{\boldsymbol{E}}^{sam}(\boldsymbol{r},-\omega))$$

i.e., the DC term and two mirror terms which can be separated into the spectral domain by using the super-gaussian spectral filter, then the second component can be obtained as:

$$\tilde{S}(\boldsymbol{r},\omega) = (\tilde{\boldsymbol{E}}^{ref}(\boldsymbol{r},\omega))^* \cdot (\tilde{\boldsymbol{E}}^{sam}(\boldsymbol{r},\omega)) \quad \text{(S10)}$$

Therefore, by rotating the polarizer, the horizontal and vertical channels of the signals can be detected by the CCD as:

$$\begin{cases} \tilde{S}_x(\boldsymbol{r},\omega) = (\tilde{E}_x^{ref}(\boldsymbol{r},\omega))^*(\tilde{E}_x^{sam}(\boldsymbol{r},\omega))e^{i\varphi_{x0}} \\ \tilde{S}_y(\boldsymbol{r},\omega) = (\tilde{E}_y^{ref}(\boldsymbol{r},\omega))^*(\tilde{E}_y^{sam}(\boldsymbol{r},\omega))e^{i\varphi_{y0}} \end{cases} \quad \text{(S11)}$$

where $\varphi_{x0}$ and $\varphi_{y0}$ are the random phase corresponding to two independent polarization channels. This equation can be rewritten as:

$$\begin{cases} \tilde{E}_x^{sam}(\boldsymbol{r},\omega) = \tilde{S}_x(\boldsymbol{r},\omega)e^{-i\varphi_{x0}}/(\tilde{E}_x^{ref}(\boldsymbol{r},\omega))^* \\ \tilde{E}_y^{sam}(\boldsymbol{r},\omega) = \tilde{S}_y(\boldsymbol{r},\omega)e^{-i\varphi_{y0}}/(\tilde{E}_y^{ref}(\boldsymbol{r},\omega))^* \end{cases} \quad \text{(S12)}$$

The time domain electric field of the sample can be given by:

$$\begin{cases} E_x^{sam}(\boldsymbol{r},t) = in\mathcal{F}[\tilde{S}_x(\boldsymbol{r},\omega)e^{-i\varphi_{x0}}/\left(\tilde{E}_x^{ref}(\boldsymbol{r},\omega)\right)^*] \\ E_y^{sam}(\boldsymbol{r},t) = in\mathcal{F}[\tilde{S}_y(\boldsymbol{r},\omega)e^{-i\varphi_{y0}}/\left(\tilde{E}_y^{ref}(\boldsymbol{r},\omega)\right)^*] \end{cases} \quad \text{(S13)}$$

where symbol "$in\mathcal{F}$" denotes the inverse Fourier transformation. Therefore, providing a homogenous reference beam across the whole detection plane of the CCD, the knowledges the spectrum and phase by using the optical spectrum analyzer and FROG (laboratory-built, based on a BBO crystal), the whole spectral domain information of the sample pulse can be obtain by using the Eq. S12. The temporal structure of the pulse can be rebuild based on the Eq. S13 after the inverse Fourier transformation.

**I.2. The experimental 2D interferograms recorded using the polarization-sensitive time-scanning MZI at a single time delay**

For the space–time pulses characterizations, Figure S4 demonstrates the 2D interferograms recorded with the polarization-sensitive time-scanning MZI for different HOP$_{SS}$ pulses. In Figure S4A, the red circles denote the well-defined singularities of the north and south pole states. It be found that the north pole state $(0,\pi/2)$ has the helicity of $-1$, and the south pole state $(0, -\pi/2)$ has the helicity of $+1$. Furthermore, the state on the Equator, the radial, azimuthal, the state for the $(\pi/2,0)$ and $(3\pi/2,0)$ (shown in Figure S4B and Figure S4C) keep a homogeneous interference patterns which coincide with the principle of the Q-plate. The inhomogeneous intensity distribution on the 2D interferograms are produced by the special shape of the donut-like pulse as well as the sectioning procedure of the time scanning techniques, whose details can be found in the 3D interferograms of Figure 5.

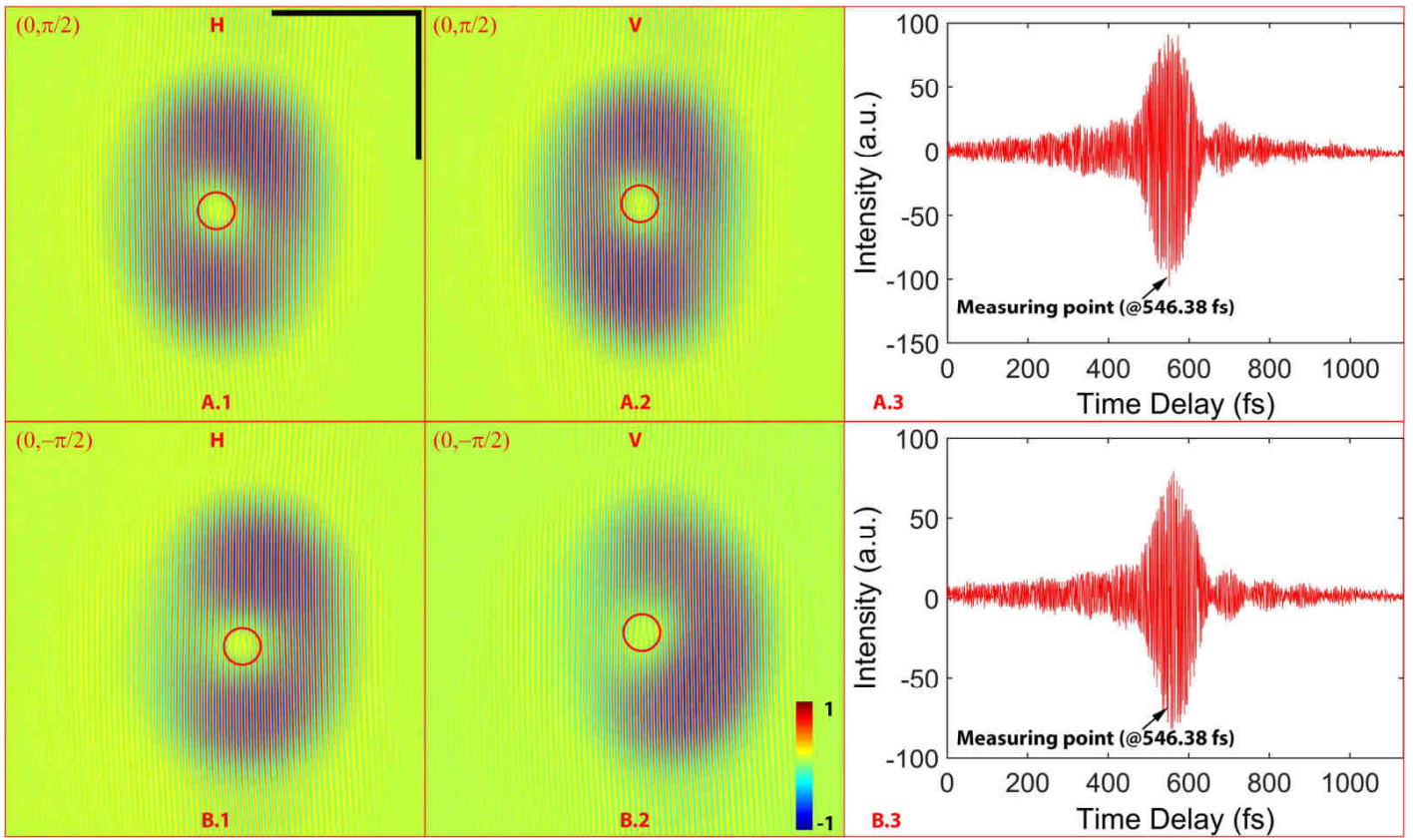

**Figure S4A.** The experimental 2D interferograms recorded with the polarization sensitive time scanning MZI for different HOP$_{SS}$ pulses. **A**: 2D interferograms for north pole state at time delay between the two interferometer's arms of 546.38 fs. **A.1** and **A.2** corresponding to the *H* and *V* channels, respectively. **A.3** is single-pixel cross-correlation signal obtained from the time scanning procedure which containing 1700 steps. **B**: 2D interferograms for south pole state at time delay of 546.38 fs. The red circle denotes the well-defined singularities of the HOP$_{SS}$ pulses. Scale bars represent 1 mm.

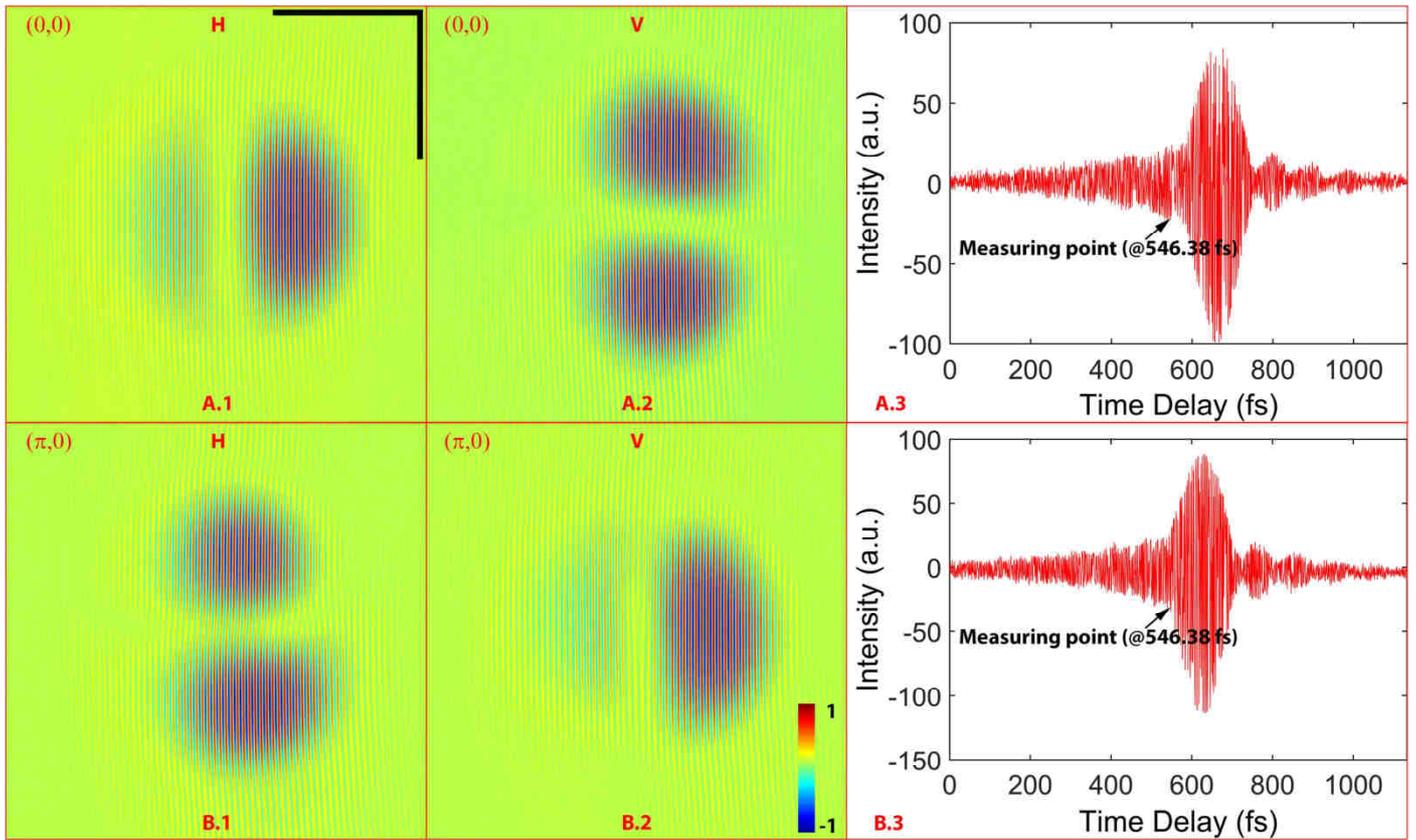

**Figure S4B.** The experimental 2D interferogram results. **A**: 2D interferograms of the radial state at time delay between the two interferometer's arms of 546.38 fs. **A.1** and **A.2** corresponding to the *H* and *V* channels, respectively. **A.3** is single-pixel cross-correlation signal obtained from the time scanning procedure. **B**: 2D interferograms of the azimuthal state at time delay of 546.38 fs. Scale bars represent 1 mm.

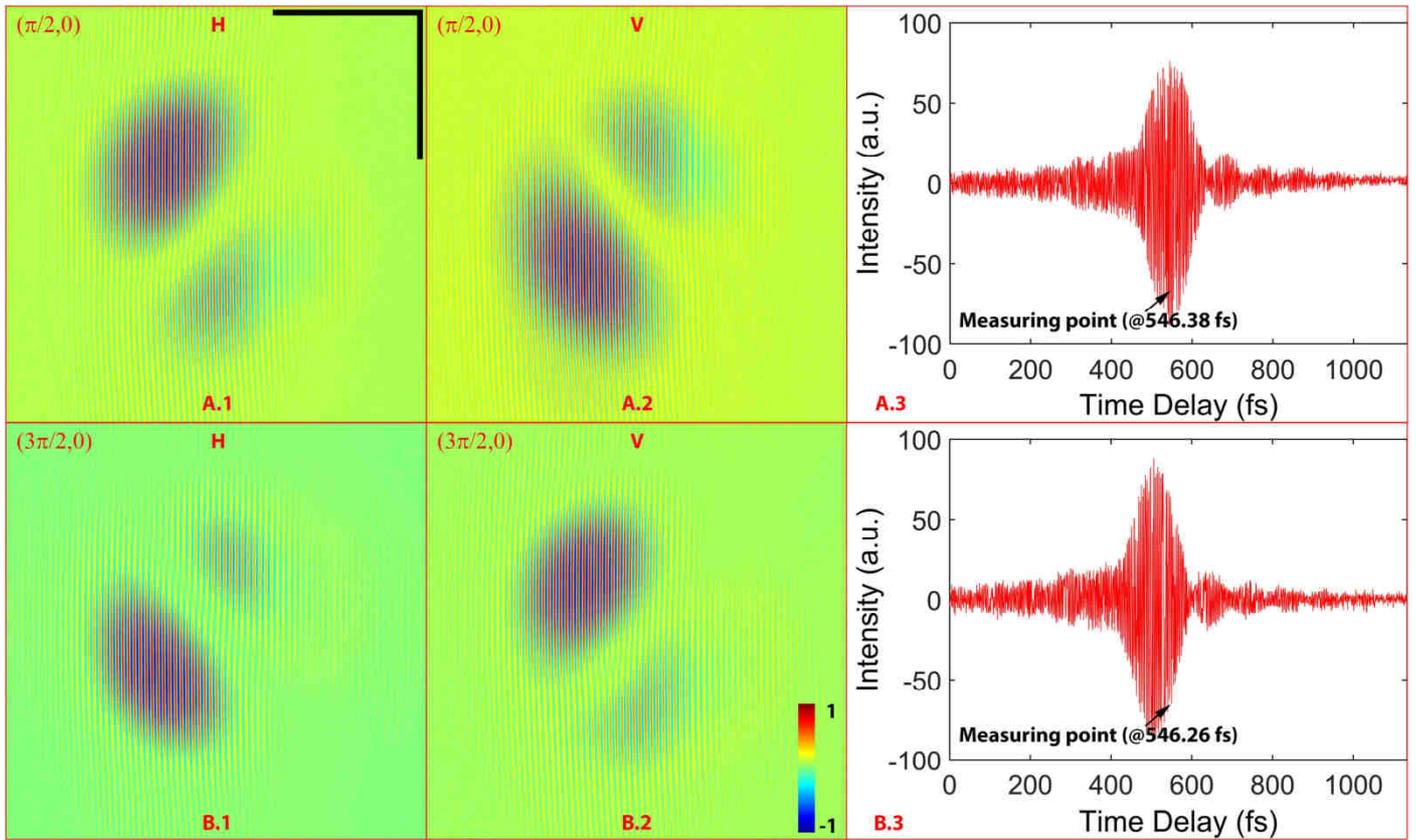

**Figure S4C.** The experimental 2D interferogram results. **A**: 2D interferograms for $(\pi/2,0)$ state at time delay between the two interferometer's arms of 546.38 fs. **A.1** and **A.2** corresponding to the *H* and *V* channels, respectively. **A.3** is single-pixel cross-correlation signal obtained from the time scanning procedure. **B**: 2D interferograms for $(3\pi/2,0)$ state at time delay of 546.26 fs. Scale bars represent 1 mm.

## I.3. The additional spectral characterization of the femtosecond pulse with HOP$_{SS}$

Figures S5A-E demonstrate additional polarization-sensitive spatial-temporal characterization of the femtosecond pulse with the states of south pole (0,-π/2), radial (0,0), azimuthal (π,0), (π/2,0) and (3π/2,0) of the HOP sphere.

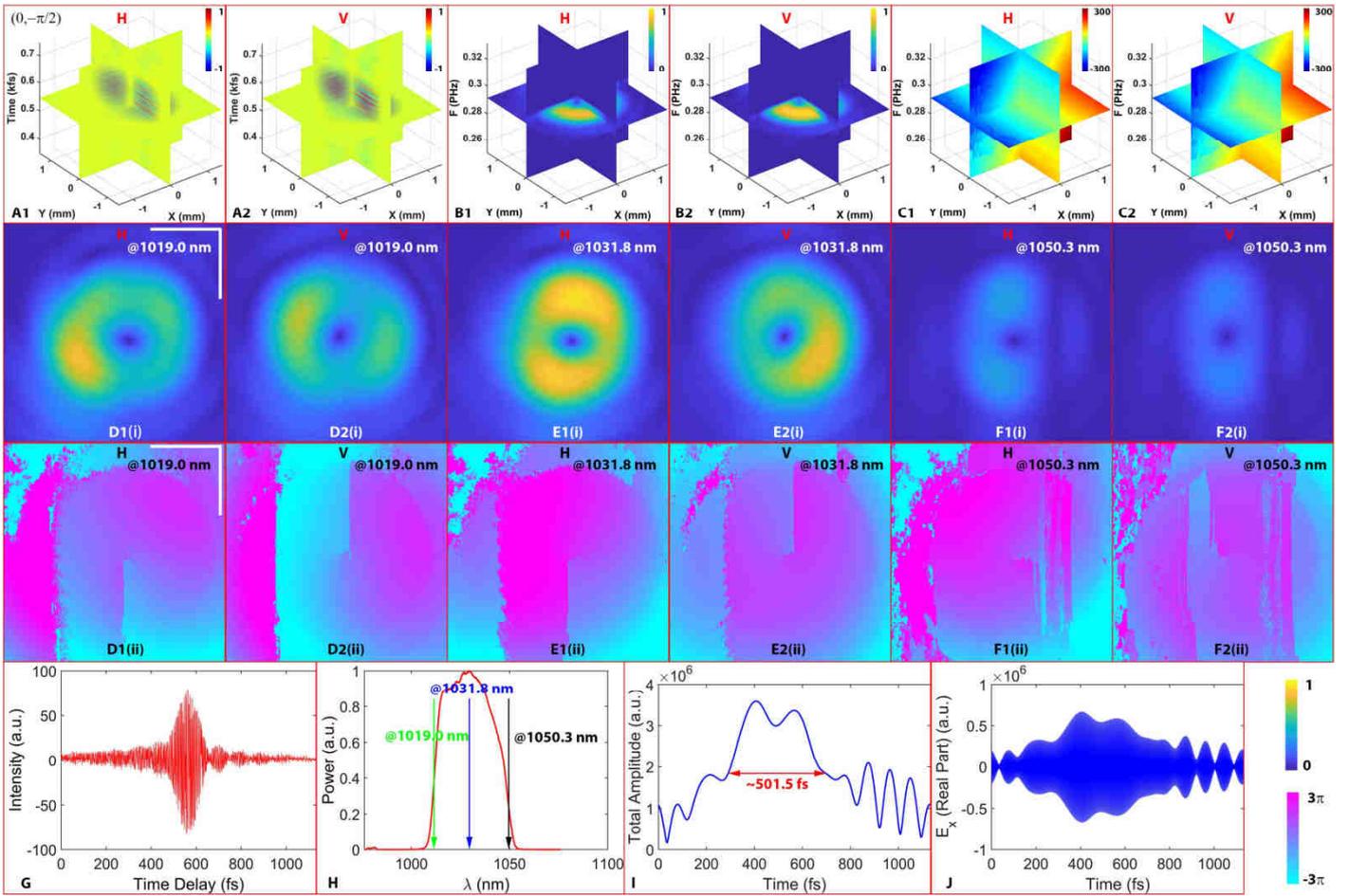

**Figure S5A.** The polarization-sensitive spatiotemporal characterizations of the pulse with south pole state $(0, -\pi/2)$. **A1-2**: the 3D interferograms of the horizontal *H* and *V* channels, respectively; **B1-2**: the corresponding 3D spectral amplitude of the two channels; **C1-2**: the corresponding 3D spectral phase. **D, E,** and **F**: the 2D amplitude (Roman numeral i) and unwrapped phase profiles (Roman numeral ii) at 1019.0 nm, 1031.8 nm and 1050.3 nm, respectively. **G**: single-pixel cross-correlation signal obtained from the time scanning procedure for *H* channel; **H**: the single-pixel spectrum; **I**: the single-pixel profile for the total temporal amplitude envelope; **J**: the corresponding real part of the pulse. Scale bars represent 1 mm.

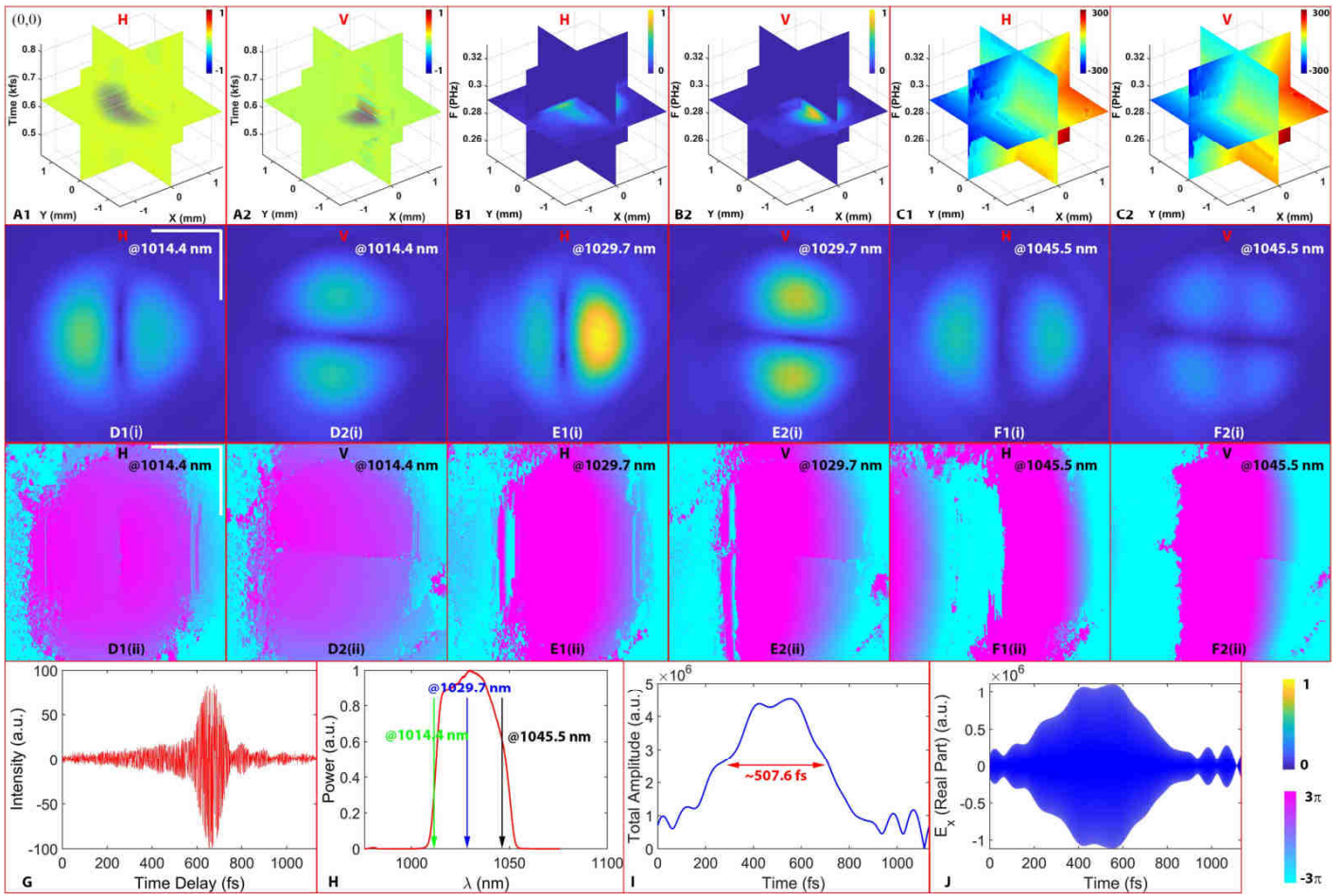

**Figure S5B.** The polarization-sensitive spatiotemporal characterization of the pulse with radial state (0,0). **A1-2**: the 3D interferograms of the *H* and vertical *V* channels, respectively; **B1-2**: the corresponding 3D spectral amplitude of the two channels; **C1-2**: the corresponding 3D spectral phase. **D, E,** and **F**: the 2D amplitude (Roman numeral i) and unwrapped phase profiles (Roman numeral ii) at 1014.4 nm, 1029.7 nm and 1045.5 nm, respectively. **G**: single-pixel cross-correlation signal obtained from the time scanning procedure for *H* channel; **H**: the single-pixel spectrum; **I**: the single-pixel profile for the total temporal amplitude envelope; **J**: the corresponding real part of the pulse. Scale bars represent 1 mm.

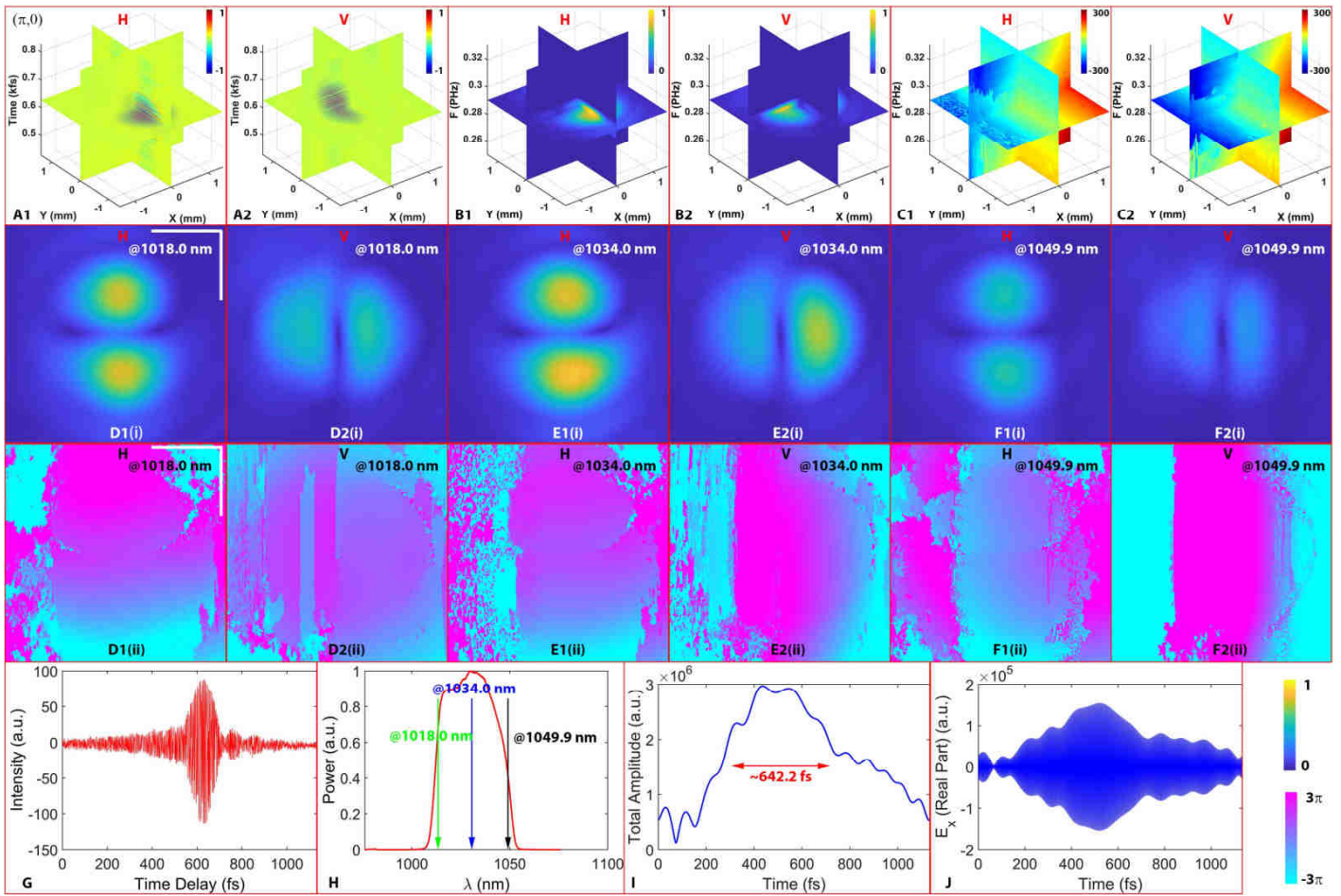

**Figure S5C.** The polarization-sensitive spatiotemporal characterizations of the pulse with azimuthal state $(\pi, 0)$. **A1-2**: the 3D interferograms of the *H* and *V* channels, respectively; **B1-2**: the corresponding 3D spectral amplitude of the two channels; **C1-2**: the corresponding 3D spectral phase. **D, E,** and **F**: the 2D amplitude (Roman numeral i) and unwrapped phase profiles (Roman numeral ii) at 1018.0 nm, 1034.0 nm and 1049.9 nm, respectively. **G**: single-pixel cross-correlation signal obtained from the time scanning procedure for *H* channel; **H**: the single-pixel spectrum; **I**: the single-pixel profile for the total temporal amplitude envelope; **J**: the corresponding real part of the pulse. Scale bars represent 1 mm.

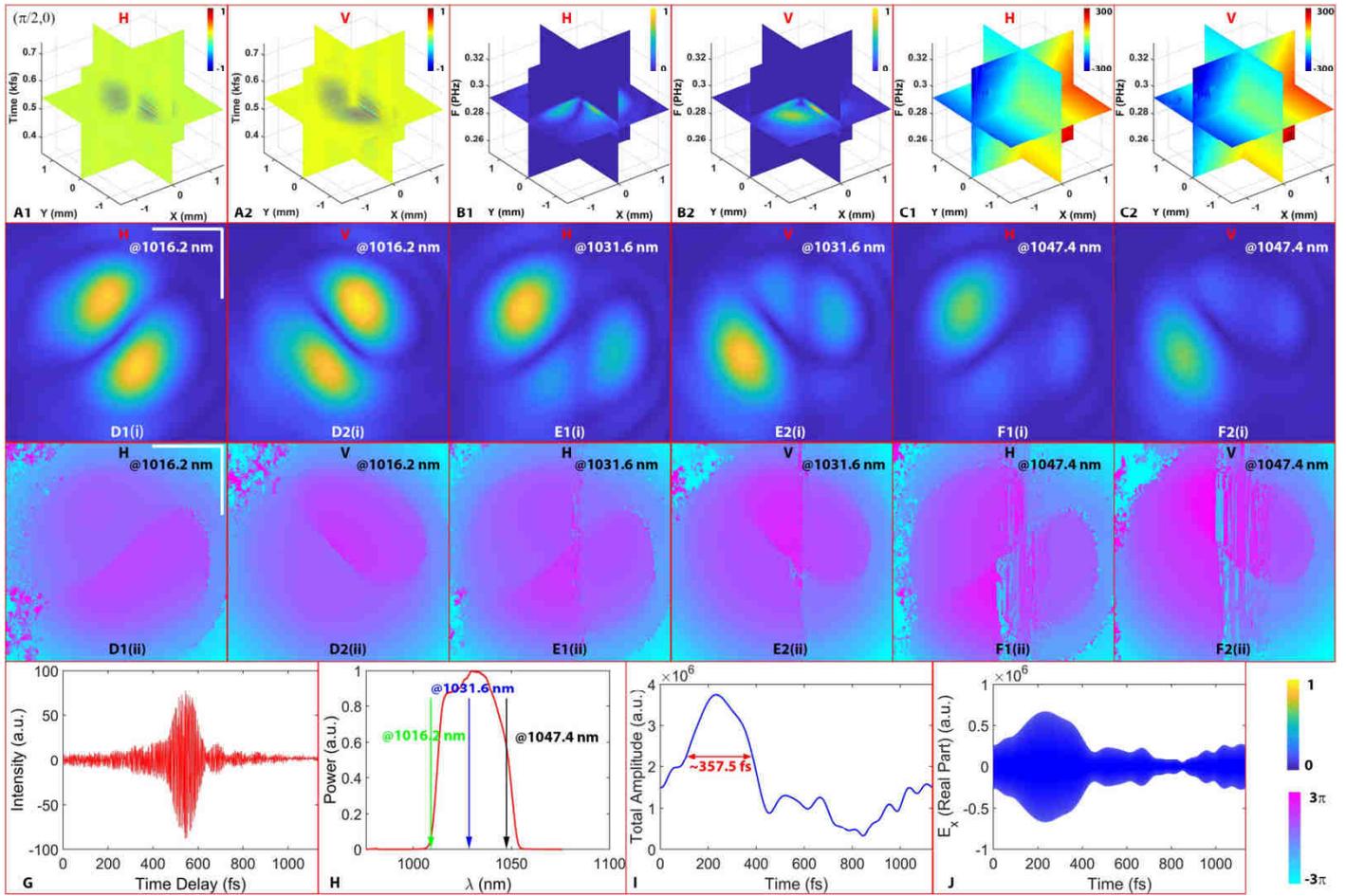

**Figure S5D.** The polarization-sensitive spatiotemporal characterizations of the pulse with ($\pi/2,0$) state of the HOP sphere. **A1-2**: the 3D interferograms of the *H* and *V* channels, respectively; **B1-2**: the corresponding 3D spectral amplitude of the two channels; **C1-2**: the corresponding 3D spectral phase. **D, E,** and **F**: the 2D amplitude (Roman numeral i) and unwrapped phase profiles (Roman numeral ii) at 1016.2 nm, 1031.6 nm and 1047.4 nm, respectively. **G**: single-pixel cross-correlation signal obtained from the time scanning procedure for *H* channel; **H**: the single-pixel spectrum; **I**: the single-pixel profile for the total temporal amplitude envelope; **J**: the corresponding real part of the pulse. Scale bars represent 1 mm.

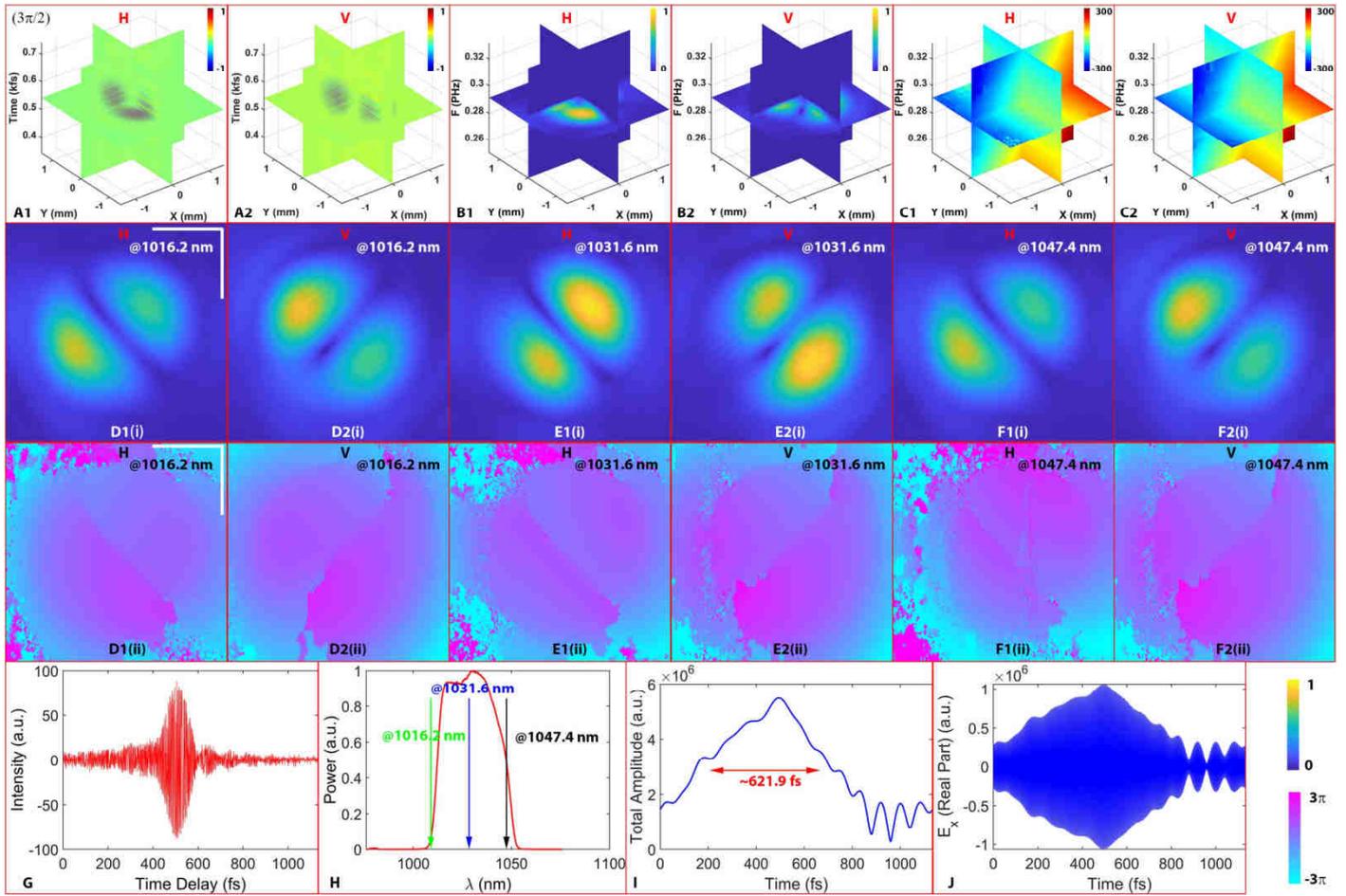

**Figure S5E.** The polarization-sensitive spatiotemporal characterizations of the pulse with ($3\pi/2$,0) state of the HOP sphere. **A1-2**: the 3D interferograms of the *H* and *V* channels, respectively; **B1-2**: the corresponding 3D spectral amplitude of the two channels; **C1-2**: the corresponding 3D spectral phase. **D, E,** and **F**: the 2D amplitude (Roman numeral i) and unwrapped phase profiles (Roman numeral ii) at 1016.2 nm, 1031.6 nm and 1047.4 nm, respectively. **G**: single-pixel cross-correlation signal obtained from the time scanning procedure for *H* channel; **H**: the single-pixel spectrum; **I**: the single-pixel profile for the total temporal amplitude envelope; **J**: the corresponding real part of the pulse. Scale bars represent 1 mm.